\documentclass[aps,prd,twocolumn,floats,superscriptaddress,showpacs,preprintnumbers,amsmath,amssymb]{revtex4}

\usepackage{amsmath,amssymb}

\usepackage{graphicx}
\usepackage{psfrag}

\usepackage{color}
\usepackage[colorlinks]{hyperref}

\newcommand{\lsim}{\mathrel{\mathop{\kern 0pt \rlap
  {\raise.2ex\hbox{$<$}}}
  \lower.9ex\hbox{\kern-.190em $\sim$}}}
\newcommand{\gsim}{\mathrel{\mathop{\kern 0pt \rlap
  {\raise.2ex\hbox{$>$}}}
  \lower.9ex\hbox{\kern-.190em $\sim$}}}

\newcommand{\beq}{\begin{equation}}
\newcommand{\eeq}{\end{equation}}
\newcommand{\bea}{\begin{eqnarray}}
\newcommand{\ena}{\end{eqnarray}}

\newcommand{\ie}{{\it i.e.}}
\newcommand{\citeeq}[1]{Eq.~(\ref{#1})}
\newcommand{\citefig}[1]{Fig.~\ref{#1}}
\newcommand{\aaa}{\hspace{0.3cm}}
\newcommand{\lD}{\mbox{$\lambda_{\rm D}$}}
\newcommand{\lDM}{\mbox{$\lambda_{\rm D}^{\rm max}$}}
\newcommand{\fluxe}{\Phi_{e^+}}

\begin{document}
\preprint{DFTT 7/2007}
\preprint{LAPTH-1187/07}
\title{Positrons from dark matter annihilation in the galactic halo: theoretical uncertainties}

\author{T. Delahaye}
\email{delahaye@lapp.in2p3.fr}
\affiliation{Laboratoire d'Annecy-le-Vieux de Physique Th\'eorique LAPTH, CNRS-SPM \\
and Universit\'e de Savoie 9, Chemin de Bellevue, B.P.110 74941 Annecy-le-Vieux, France}

\author{R. Lineros}
\email{lineros@to.infn.it}
\affiliation{Dipartimento di Fisica Teorica, Universit\`a di Torino \\
and Istituto Nazionale di Fisica Nucleare, via P. Giuria 1, I--10125 Torino, Italy}

\author{F. Donato}
\email{donato@to.infn.it}
\affiliation{Dipartimento di Fisica Teorica, Universit\`a di Torino \\
and Istituto Nazionale di Fisica Nucleare, via P. Giuria 1, I--10125 Torino, Italy}

\author{N. Fornengo}
\email{fornengo@to.infn.it}
\affiliation{Dipartimento di Fisica Teorica, Universit\`a di Torino \\
and Istituto Nazionale di Fisica Nucleare, via P. Giuria 1, I--10125 Torino, Italy}

\author{P. Salati}
\email{salati@lapp.in2p3.fr}
\affiliation{Laboratoire d'Annecy-le-Vieux de Physique Th\'eorique LAPTH, CNRS-SPM \\
and Universit\'e de Savoie 9, Chemin de Bellevue, B.P.110 74941 Annecy-le-Vieux, France}

\date{\today}

\begin{abstract}
Indirect detection signals from dark matter annihilation are studied in the positron channel. We discuss in detail
the positron propagation inside the galactic medium: we present novel solutions of the diffusion and propagation
equations and we focus on the determination of the astrophysical uncertainties which affect the positron dark matter
signal.
We find dark matter scenarios and propagation models that nicely fit existing data on the positron
fraction. Finally, we present predictions both on the positron fraction and on the flux for already
running or  planned space experiments, concluding that they have the potential to discriminate a
possible signal from the background and, in some cases, to distinguish among different astrophysical
propagation models.

\end{abstract}


\maketitle

\section{Introduction}
\label{sect:intro}

The quest for the identification of dark matter (DM), together
with the comprehension of the nature of dark energy, is one of the
most challenging problems in the understanding of the physical
world. It is therefore of utmost importance to address the problem
of the detection of the astronomical DM with different techniques
and in different channels: in underground laboratories, in
neutrino telescopes, in large--area surface detectors as well as
in space. Many efforts in both direct and indirect DM detection
have been done in the last decade, and major breakthroughs are
expected in the following years from the underground facilities
and antimatter searches in space. In the same period, the LHC will
provide crucial information on possible extensions of the Standard
Model of particle physics, where the most viable DM candidates are
predicted. We therefore are faced with the quest of signal
predictions as detailed as possible, accompanied by a realistic
estimation of their uncertainties.
\\
This paper deals with  the indirect detection of DM through positrons from the
DM pair annihilation inside the galactic halo.  Secondary positrons and
electrons are produced in the Galaxy from the collisions of cosmic-ray proton
and helium nuclei on the interstellar medium \cite{strongmoska} and are an
important tool for the comprehension of cosmic-ray propagation. Data on the
cosmic positron flux (often reported in terms of the positron fraction) have
been collected by several experiments \cite{heat,
ams,ams01,ams02,2000ApJ...532..653B,2002A&A...392..287G}. In particular, the HEAT data
\cite{heat} mildly indicate a possible excess of the positron fraction (see Eq.
\ref{fraction}) for energies above 10~GeV and with respect to the available
calculations for the secondary component \cite{strongmoska}. Different
astrophysical contributions to the positron fraction in the 10 GeV region have
been explored \cite{heat}, but only more accurate and energy extended data could
shed light on the effective presence of a bump in the positron fraction and on
its physical interpretation. Alternatively, it has been conjectured that the
possible excess of positrons found in the HEAT data could be due to the presence
of DM annihilation in the galactic halo \cite{baltz_edsjo99,Hooper:2004bq}. This
interpretation, though very exciting, is at some point limited by the
uncertainties in the halo structure and in the cosmic ray propagation modeling.
Recently, it has been shown that the boost factor due to substructures in the DM
halo depends on the positron energy and on the statistical properties of the DM
distribution \cite{Lavalle:2006vb}. In addition, it has been pointed out  that
its numerical values is quite modest \cite{Lavalle:1900wn}.

The present work  is about the issue of the propagation of primary
positrons. We inspect the full solution of the diffusion equation
in a two---zone model already tested on  several stable
and radioactive species \cite{parfit}  and quantify the
uncertainties due to propagation models, in connection with the
positron production modes.  Our results will be applied to
experiments such as PAMELA and AMS-02, which are expected to bring
a breakthrough in the cosmic antimatter searches  and in the
understanding of the positron component. In Sect. \ref{sect:2} we
present the solutions to the diffusion equation with both the
Green function formalism and the Bessel method, with a source term
due to the pair annihilations of DM particles. We introduce the
diffusive halo function, the integral on the diffusive zone
encoding the information relevant to cosmic ray propagation
through its fundamental parameters. In Sect. \ref{sect:3} we
evaluate the uncertainties due to propagation models on the
diffusive halo function, discussing the physical properties of the
propagation parameter configurations giving the extremes of the
uncertainty bands. The positron fluxes and the relevant positron
fraction are presented in Sect. \ref{sect:4}, where we compare our
results to existing data and elaborate predictions for present
running or planned experiments in space. In Sect.
\ref{sect:conclusions} we draw our conclusions.

\section{The diffusion equation and its solutions}
\label{sect:2}
%
The propagation of positrons in the galactic medium is governed by
the transport equation
\begin{equation}
{\displaystyle \frac{\partial \psi}{\partial t}} \; -
{\mathbf \nabla} \! \cdot
\left\{ K \! \left( {\mathbf x} , E \right) \, {\mathbf \nabla} \psi \right\} \; -
{\displaystyle \frac{\partial}{\partial E}}
\left\{ b(E) \, \psi \right\} = q \left( {\mathbf x} , E \right) \; ,
\label{master_2}
\end{equation}
where $\psi({\mathbf x} , E)$ denotes the positron number density
per unit energy and $q({\mathbf x} , E)$ is the positron source
term. The transport through the magnetic turbulences is described
by the space independent diffusion coefficient $K(\mathbf{x},E) \!
= \! K_0 \, \epsilon^{\delta}$ where $\epsilon \! = \! {E}/{E_0}$
and $E_{0} \! = \! 1$ GeV.
Positrons lose energy through synchrotron radiation and inverse
Compton scattering on the cosmic microwave background radiation
and on the galactic starlight at a rate $b(E) \! = \! {E_0} \,
{\epsilon^2} / {\tau_E}$ where $\tau_E \! = \! 10^{16}$~s.
%
The diffusive halo inside which cosmic rays propagate before
escaping into the intergalactic medium is pictured as a flat
cylinder with radius $R_{\rm gal} = 20$ kpc and extends along the
vertical direction from $z = - L$ up to $z = + L$. The gaseous
disk lies in the middle at $z = 0$ and contains the interstellar
material on which most of the cosmic ray spallations take place.
The half-thickness $L$ is not constrained by the measurements of
the boron to carbon ratio cosmic ray fluxes B/C. Its value could
be anywhere in the interval between 1 and 15 kpc. As cosmic rays
escape from that diffusive zone (DZ) and become scarce in the
intergalactic medium, the density $\psi$ is generally assumed to
vanish at the radial boundaries $r = R_{\rm gal}$ and $z = \pm L$.
%
Assuming steady state, the master equation~(\ref{master_2})
simplifies into
\begin{equation}
K_{0} \, \epsilon^{\delta} \, \Delta \psi \; + \;
{\displaystyle \frac{\partial}{\partial \epsilon}}
\left\{
{\displaystyle \frac{\epsilon^{2}}{\tau_{E}}} \, \psi
\right\} \; + \; q = 0 \;\; ,
\label{master_3}
\end{equation}
and may be solved by translating \cite{baltz_edsjo99} the energy
$\epsilon$ into the pseudo-time
\begin{equation}
\tilde{t}(\epsilon) = \tau_{E} \;
\left\{
v(\epsilon) = {\displaystyle \frac{\epsilon^{\delta - 1}}{1 - \delta}}
\right\} \;\; .
\label{connection_E_pseudo_t}
\end{equation}
In this formalism, the energy losses experienced by positrons are
described as an evolution in the pseudo-time $\tilde{t}$. As a
consequence, the propagation relation~(\ref{master_3}) simplifies
into the heat equation
\begin{equation}
{\displaystyle \frac{\partial \tilde{\psi}}{\partial \tilde{t}}} \; - \;
K_{0} \, \Delta \tilde{\psi} =
\tilde{q} \left( {\mathbf x} , \tilde{t} \, \right) \;\; ,
\label{master_4}
\end{equation}
where the space and energy positron density is now
$\tilde{\psi} = \epsilon^{2} \, \psi$ whereas the positron production rate
has become
$\tilde{q} = \epsilon^{2 - \delta} \, q$.

%
\begin{widetext}
In the Green function formalism, $G_{e^{+}} \left( \mathbf{x} , E
\leftarrow \mathbf{x}_{S} , E_{S} \right)$ stands for the
probability for a positron injected at $\mathbf{x}_{S}$ with the
energy $E_{S}$ to reach the location $\mathbf{x}$ with the
degraded energy $E \leq E_{S}$, and the positron density is given
by the convolution \beq \psi \left( \mathbf{x} , E \right) =
{\displaystyle \int_{E_{S} = E}^{E_{S} = + \infty}} dE_{S} \;
{\displaystyle \int_{\rm DZ}} d^{3} \mathbf{x}_{S} \,\, G_{e^{+}}
\left( \mathbf{x} , E \leftarrow \mathbf{x}_{S} , E_{S} \right)
\,\, q \left( \mathbf{x}_{S} , E_{S} \, \right) \;\; .
\label{integral_green} \eeq
\end{widetext}
In the pseudo-time approach, the positron propagator may be expressed as
\begin{equation}
G_{e^+} \left( \mathbf{x} , E \leftarrow \mathbf{x}_{S} , E_{S} \right) =
{\displaystyle \frac{\tau_{E}}{E_{0} \, \epsilon^{2}}} \;
\tilde{G} \left( \mathbf{x} , \tilde{t} \leftarrow \mathbf{x}_{S} , \tilde{t}_{S} \right)
\;\; ,
\label{positron_propagator}
\end{equation}
where $\tilde{G}$ is the Green function associated to the heat
equation~(\ref{master_4}). Without any boundary condition, this
heat propagator would be given by the 3D expression
\begin{equation}
\tilde{G} \left( \mathbf{x} , \tilde{t} \leftarrow \mathbf{x}_{S}
, \tilde{t}_{S} \right) = \left\{\!\frac{1}{ 4 \, \pi \, K_{0} \,
\tilde{\tau} \, }\!\right\}^{3/2} \! \! \! \! \exp \left\{ - \,
{\displaystyle \frac{(\Delta {\mathbf{x}})^{2}}{4 \, K_{0} \,
\tilde{\tau} \,}} \right\} , \label{propagator_reduced_3D}
\end{equation}
where $\tilde{\tau} = \tilde{t} - \tilde{t}_{S}$ is the typical
time including the diffusion process during which the positron
energy decreases from $E_{S}$ to $E$. The distance between the
source $\mathbf{x}_{S}$ and the observer $\mathbf{x}$ is $\Delta
{\mathbf{x}}$ whereas the typical diffusion length associated to
$\tilde{\tau}$ is $\lD = \sqrt{4 K_{0} \tilde{\tau}}$.
In order to implement the vertical boundary conditions $\psi ( \pm
L ) = 0$, two approaches have been so far available.

\vskip 0.1cm \noindent {\bf (i)} In the regime where the diffusion
length $\lD$ is small with respect to the DZ half-thickness $L$,
the method of the so-called electrical images consists in
implementing \cite{baltz_edsjo99} an infinite series over the
multiple reflections of the source as given by the vertical
boundaries at $+L$ and $-L$.

\vskip 0.1cm \noindent {\bf (ii)} In the opposite regime, a large
number of images needs to be considered and the convergence of the
series is a problem. Fortunately, the diffusion equation along the
vertical axis boils down to the Schr\"{o}dinger equation --
written in imaginary time -- that accounts for the behaviour of a
particle inside an infinitely deep 1D potential well that extends
from $z = - L$ to $z = + L$. The solution may be expanded as a
series over the eigenstates of the corresponding Hamiltonian
\cite{Lavalle:2006vb}.

\vskip 0.1cm \noindent None of those methods deal with the radial
boundaries at $r = R_{\rm gal}$. The diffusive halo is here a mere
infinite slab and not a flat cylinder. The Bessel approach which
we present next remedies that problem and is an improvement with
respect to the former Green formalism.

%
%
\vskip 0.5cm
\subsection{The Bessel solution}
\label{sec:bessel}

As the DZ is axisymmetric and since we will consider spherically
symmetric source terms only, we may expand the cosmic ray density
$\psi(r , z, \epsilon)$ as the Bessel series \beq \psi(r , z ,
\epsilon) = {\displaystyle \sum_{i = 1}^{\infty}} \; P_{i}(z ,
\epsilon) \; J_{0}(\alpha_{i} r / R_{\rm gal}) \;\; .
\label{bessel_psi_1} \eeq Because the $\alpha_{i}$'s are the zeros
of the Bessel function $J_{0}$, the cosmic ray density $\psi$
systematically vanishes at the radial boundaries $r = R_{\rm
gal}$. The Bessel transforms $P_{i}(z , \epsilon)$ fulfill the
diffusion equation \beq K \, \partial_{z}^{2} P_{i} \, - \, K \,
{\displaystyle \frac{\alpha_{i}^{2}}{R_{\rm gal}^{2}}} \, P_{i} \,
+ \, {\displaystyle \frac{1}{\tau_{E}}} \, \partial_{\epsilon}
\left\{ \epsilon^{2} P_{i} \right\} \, + \, Q_{i}(z , \epsilon) =
0. \label{diffusion_bessel} \eeq
\begin{widetext}
\noindent The Bessel transform $Q_{i}$ of the source distribution
$q$ is given by the usual expression \beq Q_{i}(z , \epsilon) =
{\displaystyle \frac{2}{R_{\rm gal}^{2}}} \; {\displaystyle
\frac{1}{J_{1}^{2}(\alpha_{i})}} \; {\displaystyle
\int_{0}^{R_{\rm gal}}} \, J_{0}(\alpha_{i} r / R_{\rm gal}) \,
q(r , z, \epsilon) \, r \, dr. \label{integral_bessel_radial} \eeq
\end{widetext}
Each Bessel transform $P_{i}(z , \epsilon)$ has to vanish at the
boundaries $z = - L$ and $z = + L$ and may take any value in
between. It can be therefore expanded as a Fourier series
involving the basis of functions \beq \varphi_{n}(z) = \sin ( n \,
k_{0} \, z' ) \;\; , \eeq where $k_{0} = \pi / 2 L$ and $z' = z +
L$. In our case, the DM distribution is symmetric with respect to
the galactic plane and we can restrict ourselves to the functions
$\varphi_{n}(z)$ with odd $n = 2 m + 1$ \beq \varphi_{n}(z) =
(-1)^{m} \, \cos ( n \, k_{0} \, z ) \;\; . \eeq The Bessel
transform $P_{i}(z , \epsilon)$ is Fourier expanded as \beq
P_{i}(z , \epsilon) = {\displaystyle \sum_{n = 1}^{\infty}} \;
P_{i,n}(\epsilon) \; \varphi_{n}(z) \;\; , \eeq and the same
expression holds for $Q_{i}(z , \epsilon)$ for which we need to
calculate explicitly the Fourier coefficient \beq
Q_{i,n}(\epsilon) = {\displaystyle \frac{1}{L}} \; {\displaystyle
\int_{-L}^{+L}} \, \varphi_{n}(z) \; Q_{i}(z , \epsilon) \; dz
\;\; . \label{integral_bessel_vertical} \eeq
\begin{widetext}
\noindent The Fourier transform of
equation~(\ref{diffusion_bessel}) involves the energy functions
$P_{i,n}(\epsilon)$ and $Q_{i,n}(\epsilon)$ \beq -  \, K  \, n^{2}
\, k_{0}^{2} \, P_{i,n}  \, - \, K \, {\displaystyle
\frac{\alpha_{i}^{2}}{R_{\rm gal}^{2}}} \, P_{i,n} \, + \,
{\displaystyle \frac{1}{\tau_{E}}} \, \partial_{\epsilon} \left(
\epsilon^{2} P_{i,n} \right) \, + \, Q_{i,n}(z , \epsilon) = 0
\;\; . \label{diffusion_bessel_fourier} \eeq
At this stage, as for the Green approach, we can substitute the
pseudo-time $\tilde{t}$ for the energy $\epsilon$. By defining the
new functions $\tilde{P}_{i,n} = \epsilon^{2} P_{i,n}$ and
$\tilde{Q}_{i,n} = \epsilon^{2 - \delta} Q_{i,n}$, we are led to
the heat equation \beq {\displaystyle
\frac{d\tilde{P}_{i,n}}{d\tilde{t}}} \; + \; \left\{ K_{0} \left(
n^{2} k_{0}^{2} \, + \, {\displaystyle
\frac{\alpha_{i}^{2}}{R_{\rm gal}^{2}}} \right) \right\} \,
\tilde{P}_{i,n} = \tilde{Q}_{i,n} \;\; . \eeq The solution to this
ODE is straightforward \beq \tilde{P}_{i,n}(\tilde{t}) =
{\displaystyle \int_{0}^{\tilde{t}}} \,
\tilde{Q}_{i,n}(\tilde{t}_{S}) \; \exp \left\{ - \,
\tilde{C}_{i,n} \left( \tilde{t} - \tilde{t}_{S} \right) \right\}
\; d\tilde{t}_{S} \;\; . \eeq
\end{widetext}
The argument of the exponential involves the diffusion length
$\lD$ through the pseudo-time difference $\tilde{\tau} = \tilde{t}
- \tilde{t}_{S}$ as \beq \tilde{C}_{i,n} \left( \tilde{t} -
\tilde{t}_{S} \right) = \left\{ \left( {\displaystyle \frac{n
\pi}{2 L}} \right)^{2} + {\displaystyle
\frac{\alpha_{i}^{2}}{R_{\rm gal}^{2}}} \right\} \, K_{0} \,
\tilde{\tau} \;\; . \eeq The cosmic ray positron density is given
by the double expansion \beq \psi(r , z , \epsilon) =
{\displaystyle \sum_{i = 1}^{\infty}} \; {\displaystyle \sum_{n =
1}^{\infty}} \; J_{0}(\alpha_{i} r / R_{\rm gal}) \;
\varphi_{n}(z) \; {P}_{i,n}(\epsilon) \;\; , \eeq where \beq
{P}_{i,n}(\epsilon)\! =\! {\displaystyle
\frac{\tau_{E}}{\epsilon^{2}}} {\displaystyle \int_{\epsilon}^{+
\infty}} \!\!\! {Q}_{i,n}(\epsilon_{S}) \! \exp \!\left\{
-\tilde{C}_{i,n} \left( \tilde{t} - \tilde{t}_{S} \right) \right\}
d\epsilon_{S} . \eeq We eventually get the positron flux $\fluxe =
{\beta}_{e^{+}} \psi(r , z , \epsilon) / 4 \pi$ where the positron
velocity ${\beta}_{e^{+}}$ depends on the energy $\epsilon$.

%
%
\subsection{The source term for primary positrons}
\label{sec:source}

%
Let us now consider the source term $q \left( {\mathbf x} , E
\right)$ of the master equation~(\ref{master_2}). We are here
interested in primary positrons, namely the ones that are produced
by the pair annihilations of DM particles.
According to the various supersymmetric theories, the annihilation
of a DM pair leads either to the direct creation of an
electron-positron pair or to the production of many species
subsequently decaying into photons, neutrinos, hadrons and
positrons. We have considered four possible annihilation channels
which appear in any model of weakly interacting massive particles
(WIMP).
The first one is the direct production of a e$^+$e$^-$ pair and is
actually generic for theories with extra-dimensions like the UED
models \cite{LKP_model_1,LKP_model_2,UED_models}. The energy of
the positron line corresponds to the mass of the DM species.
We have alternatively considered annihilations into W$^+$W$^-$,
$\tau^+ \tau^-$ and ${\rm b}\bar{\rm b}$ pairs. These unstable
particles decay and produce showers which may contain positrons
with a continuous energy spectrum.
Whichever the annihilation channel, the source term can be
generically written as
\begin{equation}
q \left( {\mathbf x} , E \right) = \eta \; {\left\langle \sigma v
\right\rangle} \, \left\{ {\displaystyle \frac{\rho({\mathbf
x})}{m_{\chi}}} \right\}^{2} \, f(\epsilon) \;\; . \label{source}
\end{equation}
The coefficient $\eta$ is a quantum term which depends on the
particle being or not self--conjugate~: for instance, for a fermion
it equals $1/2$ or $1/4$ depending on whether the WIMP is a
Majorana or a Dirac particle. In what follows, we have considered
a Majorana type species and taken $\eta = 1/2$.
The annihilation cross section is averaged over the momenta of the
incoming DM particles to yield $\left\langle \sigma v
\right\rangle$, the value of which depends on the specific SUSY
model and is constrained by cosmology. We have actually taken here
a benchmark value of $2.1 \times 10^{-26}$ cm$^{3}$ s$^{-1}$ which
leads to a relic abundance of $\Omega_{\chi} h^{2} \sim 0.14$ (in
agreement with the WMAP observations) under the hypothesis of
dominant s--wave annihilation and by means of the relation:

\begin{equation}
\Omega h^{2} = 8.5 \cdot 10^{-11} \,\frac{g^{1/2}_{\star}(x_{f})}{g_{\star S}(x_{f})}
\,\frac{\mbox{GeV$^{-2}$}}{x_{f^{-1}}\left\langle \sigma v\right\rangle} = 
\frac{3 \cdot 10^{-27} \mbox{cm}^{3} \mbox{s}^{-1}}{\left\langle \sigma v\right\rangle}
\end{equation}
where $x_{f}=m_{\chi}/T_{f}\simeq (20\div 25)$ with $T_{f}$ the freeze--out temperature
and where
$g_{\star}(x_{f})$ and $g_{\star S}(x_{f})$ denote
the effective number of degrees of freedom
of the energy and entropy density at freeze--out, respectively.

The DM mass $m_{\chi}$ is unknown. In the case of neutralinos,
theoretical arguments as well as the LEP and WMAP results
constrain this mass to range from a few GeV \cite{Bottino:2002ry,
Bottino:2003iu,Belanger:2002nr,Hooper:2002nq} up to a few TeV.
Keeping in mind the positron HEAT excess, we have chosen a
neutralino mass of 100 GeV. We have also analyzed the positron
signal yielded by a significantly heavier DM particle of 500 GeV.
Finally, the energy distribution of the positrons produced in a
single WIMP annihilation is denoted by $f(\epsilon) \equiv dN_{\rm
e^{+}}/dE_{\rm e^{+}}$ and has been evaluated with the help of the
Pythia Monte-Carlo \cite{Pythia}.

%
The only astronomical ingredient in the source term~(\ref{source})
is the DM distribution $\rho({\mathbf x})$ inside the Milky Way
halo. We have considered the generic profile \beq \rho(r) =
\rho_{\odot} \, \left\{ {\displaystyle \frac{r_{\odot}}{r}}
\right\}^{\gamma} \, \left\{ {\displaystyle \frac{1 \, + \, \left(
r_{\odot} / r_{s} \right)^{\alpha}} {1 \, + \, \left( r / r_{s}
\right)^{\alpha}}} \right\}^{\left( \beta - \gamma \right) /
\alpha} \;\; , \label{eq:profile} \eeq where $r_{\odot} = 8.5$ kpc
is the galactocentric distance of the solar system. Notice that
$r$ denotes here the radius in spherical coordinates. The solar
neighborhood DM density has been set equal to $\rho_{\odot} = 0.3$
GeV cm$^{-3}$.
Three profiles have been discussed in this work~: an isothermal
cored distribution~\cite{bahcall} for which $r_{s}$ is the radius
of the central core, the Navarro, Frenk and White profile
\cite{nfw} (hereafter NFW) and Moore's model \cite{moore}. The NFW
and Moore profiles have been numerically established thanks to
N-body simulations. In the case of the Moore profile, the index
$\gamma$ lies between 1 and 1.5 and we have chosen a value of 1.3
-- see Tab.~\ref{tab:indices}.
%
\begin{table}[t]
\vskip 0.5cm
{\begin{tabular}{@{}|l|c|c|c|c|@{}}
\hline
Halo model & $\alpha$ & $\beta$ & $\gamma$ & $r_s$ [kpc] \\
\hline
\hline
Cored isothermal~\cite{bahcall}
& {\aaa} 2 {\aaa} & {\aaa} 2 {\aaa} & {\aaa} 0 {\aaa} & {\aaa} 5 {\aaa} \\
Navarro, Frenk \& White~\cite{nfw}
&        1        &        3        &        1        &        20       \\
Moore~\cite{moore}
&        1.5      &        3        &        1.3      &        30       \\
\hline
\end{tabular}}
\caption{
Dark matter distribution profiles in the Milky Way.
\label{tab:indices}}
\end{table}
%
The possible presence of DM substructures inside those smooth
distributions enhances the annihilation signals by the so-called
boost factor whose value is still open to debate.

%
The positron flux at the Earth may be expressed as \beq \fluxe =
\frac{{\beta}_{e^{+}}}{4 \pi} \{ \psi(\odot , \epsilon) \equiv
\kappa \, {\displaystyle \frac{\tau_{E}}{\epsilon^{2}}} \,
{\displaystyle \int_{\displaystyle \epsilon}^{+ \infty}} \!\!
d\epsilon_{S} \, f(\epsilon_{S}) \, \tilde{I}(\lD) \} ,
\label{eq:fluxe} \eeq where the information pertinent to particle
physics has been factored out in \beq \kappa = \eta \;
{\left\langle \sigma v \right\rangle} \, \left\{ {\displaystyle
\frac{\rho_{\odot}}{m_{\chi}}} \right\}^{2} \;\; . \eeq The
diffusive halo integral $\tilde{I}$ depends on the input energy
$\epsilon_{S}$ and on the observed energy $\epsilon$ through the
diffusion length $\lD$ given by \beq \lD^{2} = 4 K_{0} \tau_{E} \,
\left\{ {\displaystyle \frac{\epsilon^{\delta - 1} -
\epsilon_{S}^{\delta - 1}}{1 - \delta}} \right\} \;\; .
\label{definition_lD} \eeq In the Green formalism, the halo
function $\tilde{I}$ may be expressed as the convolution of the
reduced propagator $\tilde{G}$ -- see
Eq.~(\ref{positron_propagator}) and (\ref{propagator_reduced_3D})
-- with the DM density squared $({\rho}/{\rho_{\odot}})^{2}$ over
the diffusive zone \beq \tilde{I}(\lD) = {\displaystyle \int_{\rm
DZ}} d^{3} \mathbf{x}_{S} \; \tilde{G} \left( \odot , \epsilon
\leftarrow \mathbf{x}_{S} , \epsilon_{S} \right) \; \left\{
{\displaystyle \frac{\rho(\mathbf{x}_{S})}{\rho_{\odot}}}
\right\}^{2} \;\; . \label{I_tilde_Green} \eeq
\begin{widetext}
\noindent Alternatively, in the Bessel approach, the halo integral
$\tilde{I}$ is given by the radial and vertical expansions \beq
\tilde{I}(\lD) = {\displaystyle \sum_{i = 1}^{\infty}} \,
{\displaystyle \sum_{n = 1}^{\infty}} \,\, J_{0}(\alpha_{i} r /
R_{\rm gal}) \; \varphi_{n}(z) \; \exp \left\{ - \tilde{C}_{i,n}
\left( \tilde{t} - \tilde{t}_{S} \right) \right\} \; R_{i,n} \;\;
, \label{I_tilde_Bessel} \eeq where the coefficients $R_{i,n}$ are
the Bessel and Fourier transforms of the DM density squared
$({\rho}/{\rho_{\odot}})^{2}$.
\end{widetext}
We insist again on the fact that the true argument of the halo
function, whatever the approach followed to derive it, is the
positron diffusion length $\lD$. This integral encodes the
information relevant to cosmic ray propagation through the height
$L$ of the diffusive zone, the normalization $K_{0}$ of the
diffusion coefficient and its spectral index $\delta$. It is also
the only relevant quantity concerning the DM distribution. The
analysis of the various astrophysical uncertainties that may
affect the positron signal of annihilating WIMPs will therefore be
achieved by studying the behavior of $\tilde{I}$.
%
\begin{figure}[t] \centering
\includegraphics[width=\columnwidth]{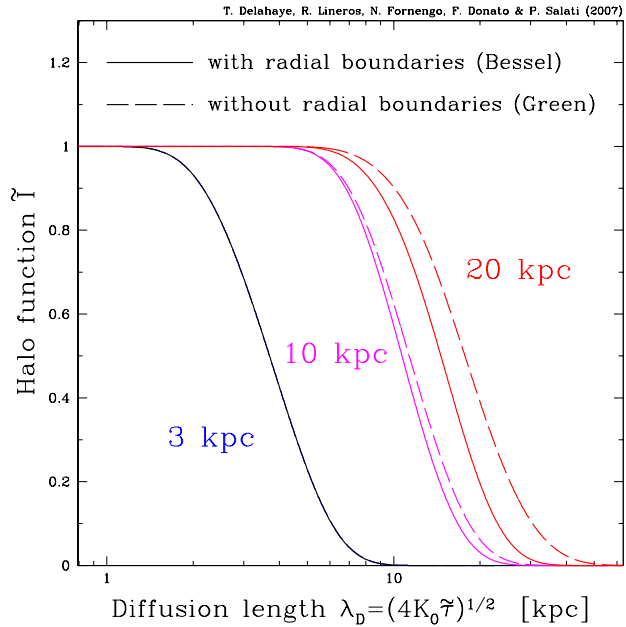}
\caption{Influence of the radial boundary condition for a slab half-thickness $L$ of 3, 10 and 20 kpc
($R_{\rm gal} = 20$ kpc).
The thicker the slab, the larger the error when neglecting the radial boundary.
On the contrary, for small values of $L$, positrons produced near the radial outskirts
of the diffusive halo escape into the intergalactic medium and do not contribute to the
signal at the Earth. Implementing correctly the radial boundary condition is not relevant
in that regime.
\label{fig1}}
\end{figure}
%

%
%
\subsection{The Bessel method versus the Green approach}

The diffusive halo integral $\tilde{I}(\lD)$ may be calculated by
using either the Bessel expansion method or the Green function
approach. We investigate here the relevance of each as a function
of the diffusion length $\lD$.

To commence, the DM distribution is taken in \citefig{fig1} to be
constant throughout the diffusive zone with $\rho = \rho_{\odot}$.
Both methods -- Green or Bessel -- do not give the same result as
soon as $L$ is large enough. Neglecting the radial boundary
condition -- the cosmic ray density vanishes at $r = R_{\rm gal}$
-- leads to overestimate the halo function $\tilde{I}$ when the
diffusion slab is thick. This can be easily understood~: if the
slab is thin enough, a positron created near the radial boundary
has a large probability to hit the vertical borders of the
diffusive zone at $\pm L$ and hereby to escape into the
intergalactic medium, never reaching the Earth. If so, the
positron horizon does not reach radially the outskirts of the
diffusive zone and the Green approach (long dashed curves)
provides a very good approximation to the correct value of
$\tilde{I}$ as given by the Bessel expansion (solid lines).
Conversely, if the slab is thick, we detect at the Earth a non
negligible fraction of positrons produced near the radial boundary
and the Green approximation is no longer acceptable. This is
particularly true for the red curves of the $L = 20$ kpc case
where the Green result largely overestimates the exact value.
This justifies the use of the Bessel expansion method which
improves upon the previous treatments of positron propagation and
is one of the novelties of this article.
%
\begin{figure*}[t]
\begin{minipage}[h!]{0.49\textwidth}
\centering
\includegraphics[width=\textwidth]{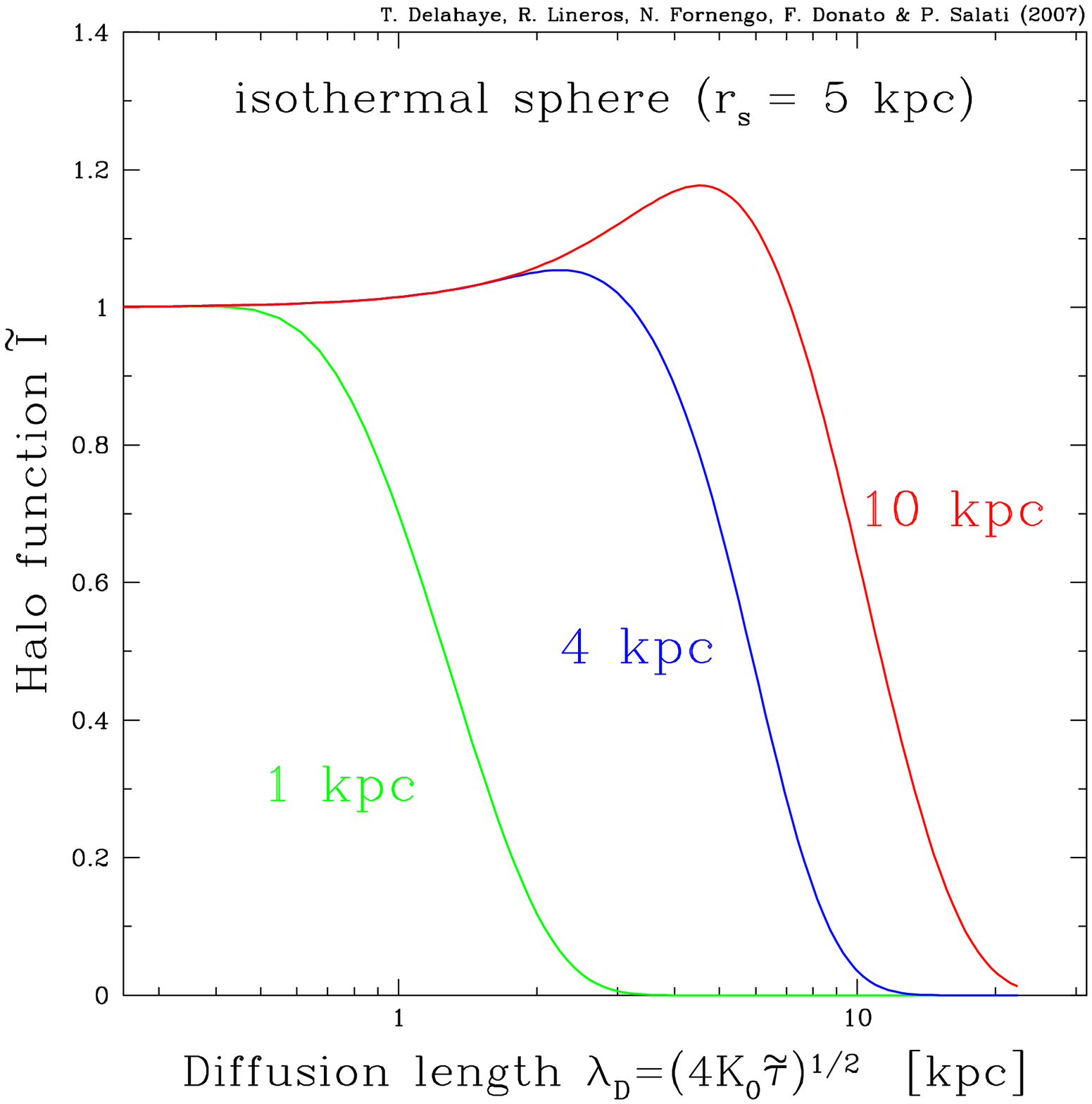}\\
\end{minipage}
\begin{minipage}[h!]{0.02\textwidth}
\end{minipage}
\begin{minipage}[h!]{0.49\textwidth}
\centering
\includegraphics[width=\textwidth]{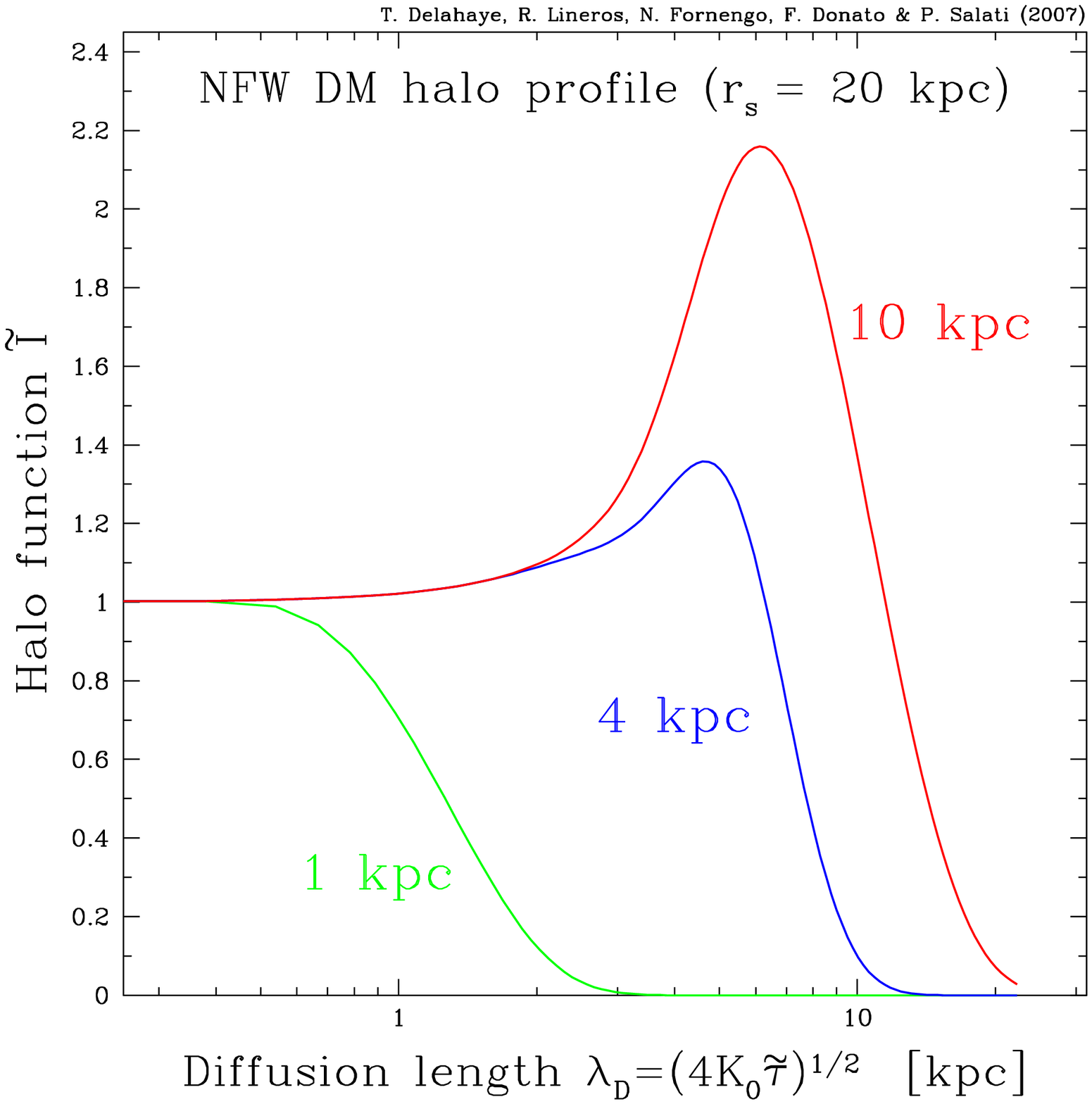}\\
\end{minipage}
\caption{ The halo convolution $\tilde{I}$ is plotted as a
function of the diffusion length $\lD$ for various values of the
slab half-thickness $L$. The left panel features the case of an
isothermal DM distribution whereas a NFW profile has been assumed
in the right panel -- see Tab.~\ref{tab:indices}. When $L$ is
large enough for the positron horizon to reach the galactic center
and its denser DM distribution, a maximum appears in the curves
for $\lD \sim r_{\odot}$. \label{fig2}}
\end{figure*}
%

As already discussed, a change in the normalization $K_{0}$ or in
the index $\delta$ of the cosmic ray diffusion coefficient leads
only to a variation of the diffusion length $\lD$ through which
those parameters appear. Notice that the relation that links the
diffusion length $\lD$ to the diffusive zone integral $\tilde{I}$
is not affected by those modifications.
On the contrary, the half-thickness $L$ of the diffusive slab has
a direct influence on the overall shape of $\tilde{I}$ as a
function of $\lD$ as is clear in \citefig{fig2}. In the left panel
an isothermal distribution has been assumed whereas the right
panel features the case of a NFW profile.
For small values of $L$ -- see the green curve for which $L = 1$
kpc -- the positron horizon is fairly limited.  Because the
positrons detected at the Earth merely originate from a very near
region, the DM profile which we probe is essentially uniform. As
in \citefig{fig1}, the DZ integral $\tilde{I}$ is unity below $\lD
\sim L$ and collapses for larger values of the diffusion length.
For a thicker slab, the cosmic ray positron flux at the Earth gets
sensitive to the center of the galaxy. That is why the halo
integral $\tilde{I}$ exhibits a maximum for a diffusion length
$\sim 5 - 7$ kpc, a value close to the galactocentric distance
$r_{\odot} = 8.5$ kpc of the solar system. In both panels, the
larger $L$, the more visible the bump. Notice also that the
steeper the DM profile, the higher the maximum.
The curves featured in \citefig{fig2} point towards the importance
of calculating correctly the influence of the DM located at the
galactic center, when $L$ is large. We therefore need to assess
the relative merits of the Bessel and Green approaches in doing
so.
%
\begin{figure*}[t]
\begin{minipage}[h!]{0.49\textwidth}
\centering
\includegraphics[width=\textwidth]{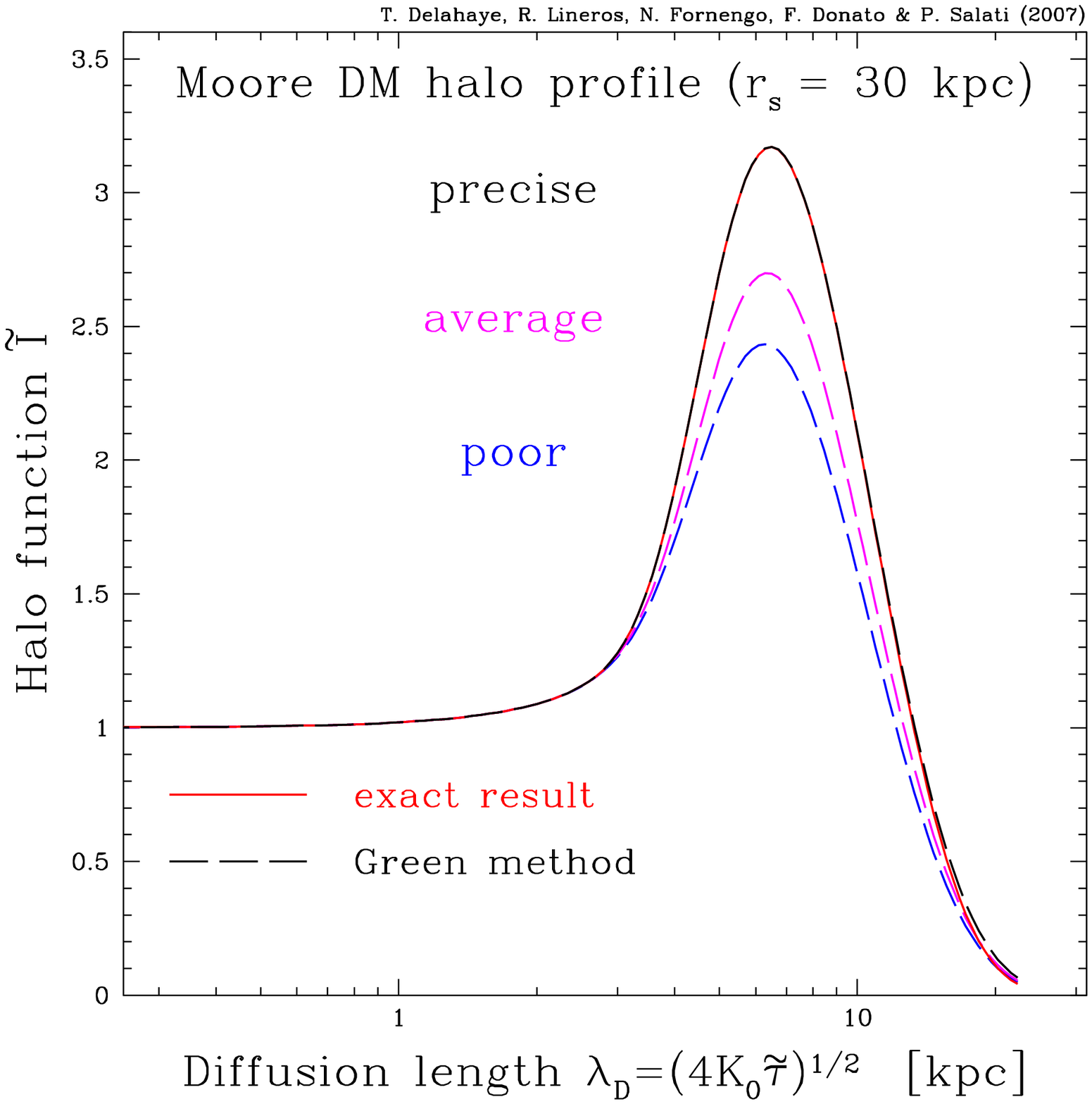}\\
\end{minipage}
\begin{minipage}[h!]{0.02\textwidth}
\end{minipage}
\begin{minipage}[h!]{0.49\textwidth}
\centering
\includegraphics[width=\textwidth]{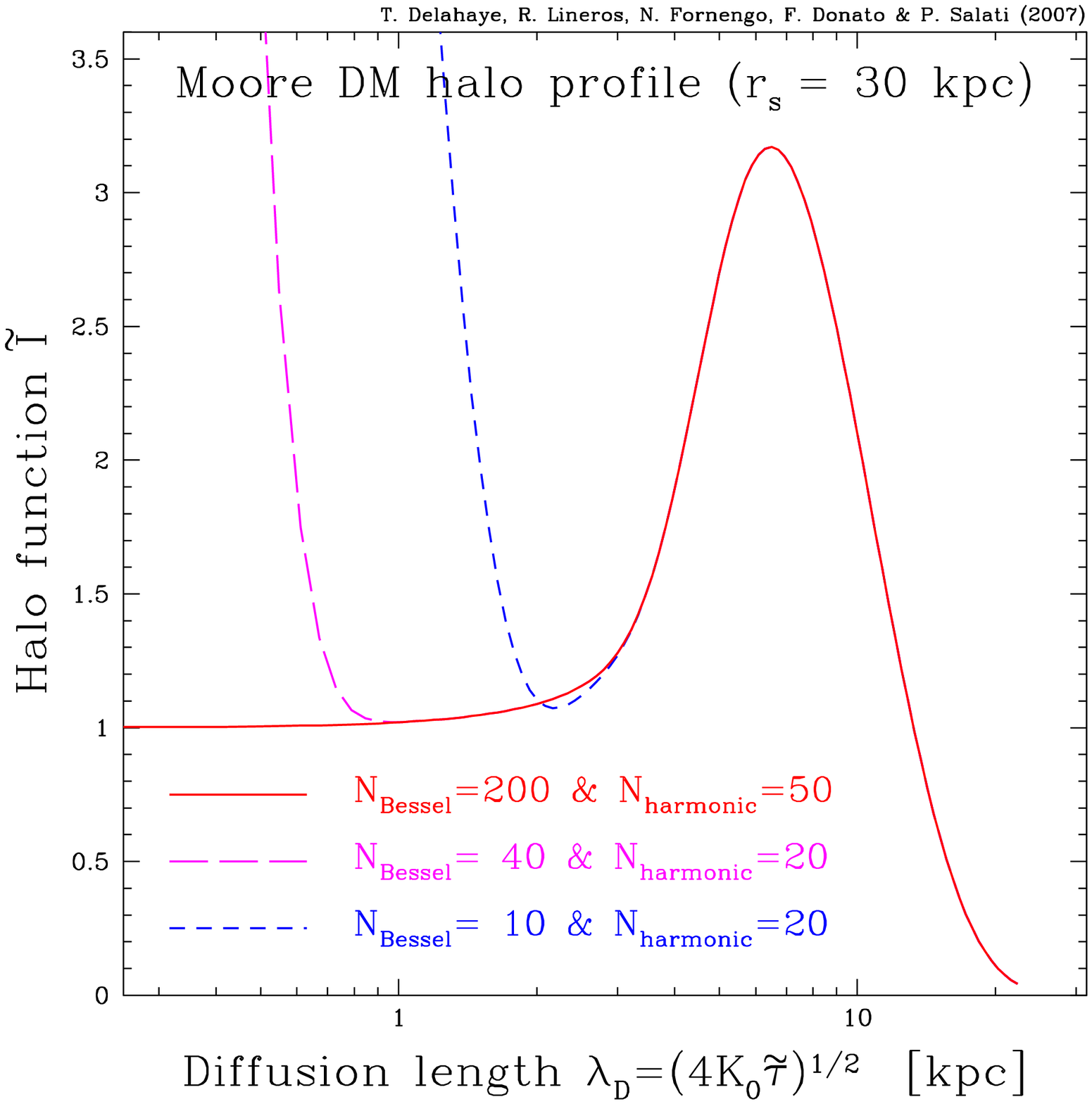}\\
\end{minipage}
\caption{
The halo integral $\tilde{I}$ is plotted as a function of the diffusion length $\lD$
in the case of a Moore profile with $L = 10$ kpc.
In the left panel, the results obtained with the Green function
method are featured by the long-dashed curves and may be compared
to the exact solution and its solid red line.
In the right panel, the numbers $N_{\rm Bessel}$ and $N_{\rm
harmonic}$ of the eigenfunctions considered in the Bessel
expansion~(\ref{I_tilde_Bessel}) have been varied. The various
curves reproduce astonishingly well the bump but diverge at small
$\lD$ when too few Bessel and Fourier terms are considered. }
\label{fig3}
\end{figure*}
%

To achieve that goal, we have selected Moore's model with a very
steep and dense DM central distribution. In the left panel of
\citefig{fig3}, the dashed curves are obtained by the Green
method. The convolution~(\ref{I_tilde_Green}) is numerically
calculated by summing over the grid of elementary cells into which
the diffusive halo has been split. The resolution of that grid
matters. For very small cells, the correct behavior of $\tilde{I}$
is recovered and the dashed black curve is superimposed on the
solid red line of the exact result derived with the Bessel
expansion technique. However, the price to pay is an unacceptable
CPU time. We had actually to break the inner 1 kpc into $8 \times
10^{5}$ cells in order for the integral~(\ref{I_tilde_Green}) to
converge. When the resolution of the grid is relaxed by increasing
the size of the Green cells, the bump is dramatically
underestimated. This is especially clear for the dashed blue curve
(labeled poor) where the DZ grid contains only a few $10^{4}$
cells. On the other hand, notice that even in that case, the
correct result is obtained for a diffusion length smaller than
$\sim 3$ kpc.
In the right panel, we have concentrated on the Bessel method and
numerically calculated the expansion~(\ref{I_tilde_Bessel}). We
have performed the summation up to a Bessel order of $N_{\rm
Bessel}$ and a Fourier order of $N_{\rm harmonic}$. The exact
result -- featured by the solid red curve -- incorporates a large
number of modes and is once again obtained at the price of a long
CPU time. If the expansion~(\ref{I_tilde_Bessel}) is truncated
earlier -- see the dashed blue line -- we observe that the correct
value of $\tilde{I}$ is completely missed when the diffusion
length is small. In that regime, the positrons that are detected
at the Earth originate mostly from the solar neighborhood. A large
number $N_{\rm Bessel}$ of radial modes is needed in order for the
Bessel transforms $R_{i,n}$ to interfere destructively with each
other so that the influence of the galactic center is erased.
For larger values of $\lD$, the exponential terms in
\citeeq{I_tilde_Bessel} force the series over $R_{i,n}$ to
converge rapidly. The exact value of the halo integral $\tilde{I}$
can be recovered even with as few terms as 10 Bessel modes and 20
Fourier harmonics. Notice how well the peak at $\lD \sim 7$ kpc is
reproduced  by all the curves, whatever $N_{\rm Bessel}$ and
$N_{\rm harmonic}$. This peak was obtained with difficulty in the
Green function approach.
We therefore strongly advise to use the Green method as long as
$\lD$ is smaller than $\sim 3$ kpc whereas the Bessel expansion
technique should be preferred above that value. This prescription
allows a fast and accurate evaluation of the halo function
$\tilde{I}$. We can safely embark on an extensive scan of the
cosmic ray propagation parameters and assess the theoretical
uncertainties that may affect the positron DM signal at the Earth.
%
\begin{figure}[t] \centering
\includegraphics[width=\columnwidth]{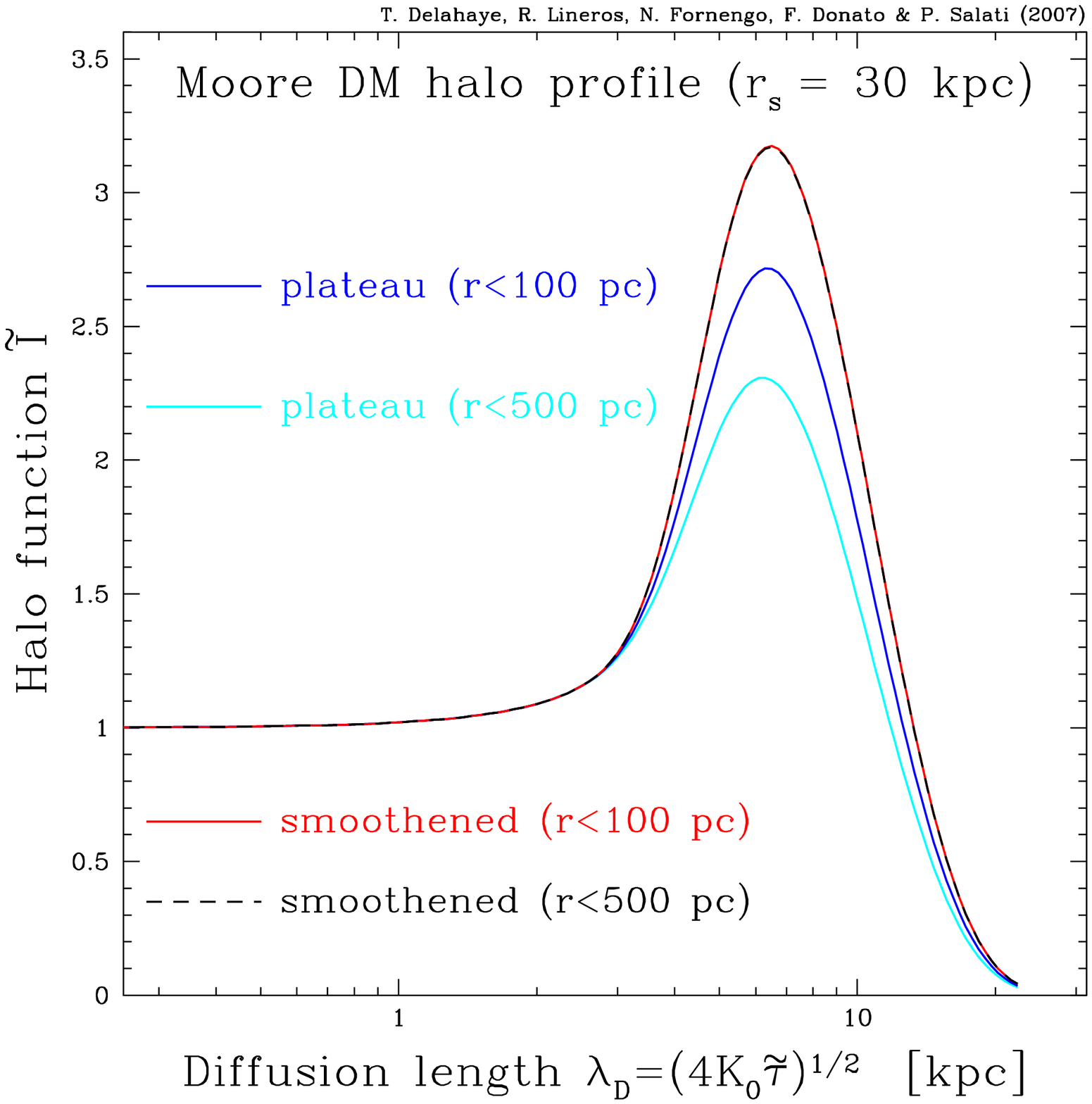}
\caption{
Same plot as before where the central DM profile within a radius $r_{0}$ is either
a plateau at constant density $\rho_{0}$ or the smooth distribution $\rho^{\ast}$
of \citeeq{profile_timur}.
In the former case, the bump which $\tilde{I}$ exhibits is significantly underestimated
even for values of $r_{0}$ as small as 100 pc -- solid dark blue -- and drops as
larger values are considered -- solid light blue. On the contrary, if the DM cusp
is replaced by the smooth profile $\rho^{\ast}$, the halo integral no longer depends
on the renormalization radius $r_{0}$ and the solid red and long-dashed black curves
are superimposed on each other.
\label{fig4}}
\end{figure}
%

%
%
\subsection{The central divergence}

Numerically derived DM profiles -- NFW and Moore -- exhibit a
divergence at the center of the Milky Way. The density increases
like $r^{- \gamma}$ for small radii -- see \citeeq{eq:profile} --
but cannot exceed the critical value for which the WIMP
annihilation timescale is comparable to the age of the galactic
bulge. The saturation of the density typically occurs within a
sphere of $\sim 10^{-7}$ pc, a much shorter distance than the
space increment in the numerical integrals, {\ie}, the Green
convolution~(\ref{integral_green}) and
relations~(\ref{integral_bessel_radial}) and
(\ref{integral_bessel_vertical}) for the Bessel expansion
technique.
Fortunately, this numerical difficulty can be eluded by noticing
that the Green propagator $G_{e^{+}} \left( \odot , \epsilon
\leftarrow r \sim 0 , \epsilon_{S} \right)$ which connects the
inner Galaxy to the Earth does not vary much over the central DM
distribution. This led us to replace inside a sphere of radius
$r_{0}$ the $r^{- \gamma}$ cusp with the smoother profile \beq
\rho^{\ast}(r) = \rho_{0} \, \left\{ 1 + a_{1} \, \text{sinc}(\pi
x) + a_{2} \, \text{sinc}(2 \pi x) \right\}^{1/2} \;\; ,
\label{profile_timur} \eeq where $x = r / r_{0}$ is the reduced
radius. The coefficients $\rho_{0}$, $a_{1}$ and $a_{2}$ are
obtained by requiring that both the smooth density $\rho^{\ast}$
and its first derivative ${d\rho^{\ast}}/{dr}$ are continuous at
$r_{0}$. The other crucial condition is the conservation of the
total number of annihilations within $r_{0}$ as the diverging cusp
$\rho \propto r^{- \gamma}$ is replaced by the distribution
$\rho^{\ast}$. These conditions imply that $\rho_{0} \equiv
\rho(r_{0})$ whereas
\begin{eqnarray}
a_1      &=& a_2 \, + \, 2 \, \gamma \;\; , \\
a_2      &=& 8 \, \gamma
\left\{
{\displaystyle \frac{\pi^{2} - 9 + 6 \gamma}{9(3 - 2\gamma)}}
\right\} \;\; .
\end{eqnarray}

In \citefig{fig4}, the halo integral $\tilde{I}$ is plotted as a
function of the diffusion length $\lD$ in the case of the Moore
profile and assuming a slab half-thickness $L = 10$ kpc.
Within the radius $r_{0}$, the central DM divergence has been
replaced either by a plateau with constant density $\rho_{0}$ or
by the renormalized profile~(\ref{profile_timur}).
In the case of the plateau, the maximum which $\tilde{I}$ reaches
for a diffusion length $\lD \sim 7$ kpc is underestimated even if
values as small as 100 pc are assumed for $r_{0}$. The larger that
radius, the fewer the annihilations taking place within $r_{0}$ as
compared to the Moore cusp and the worse the miscalculation of the
halo integral. Getting the correct result featured by the solid
red line would require a plateau radius so small that the CPU time
would explode.
On the contrary, we observe that the halo integral $\tilde{I}$ is
stable with respect to a change of $r_{0}$ when the renormalized
density $\rho^{\ast}$ is used.

%
\begin{figure*}[t]
\begin{minipage}[h!]{0.49\textwidth}
\centering
\includegraphics[width=\textwidth]{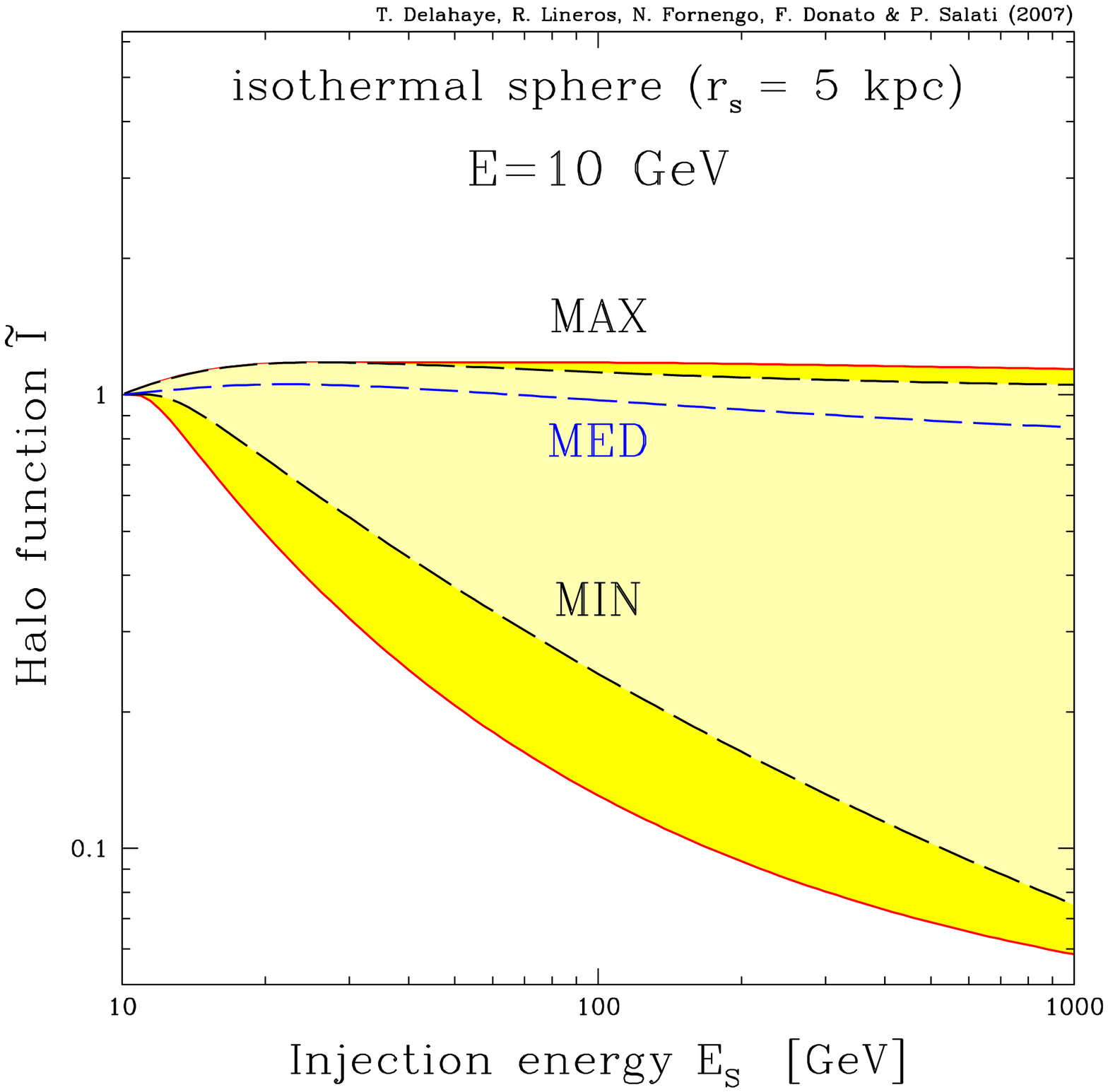}\\
\end{minipage}
\begin{minipage}[h!]{0.02\textwidth}
\end{minipage}
\begin{minipage}[h!]{0.49\textwidth}
\centering
\includegraphics[width=\textwidth]{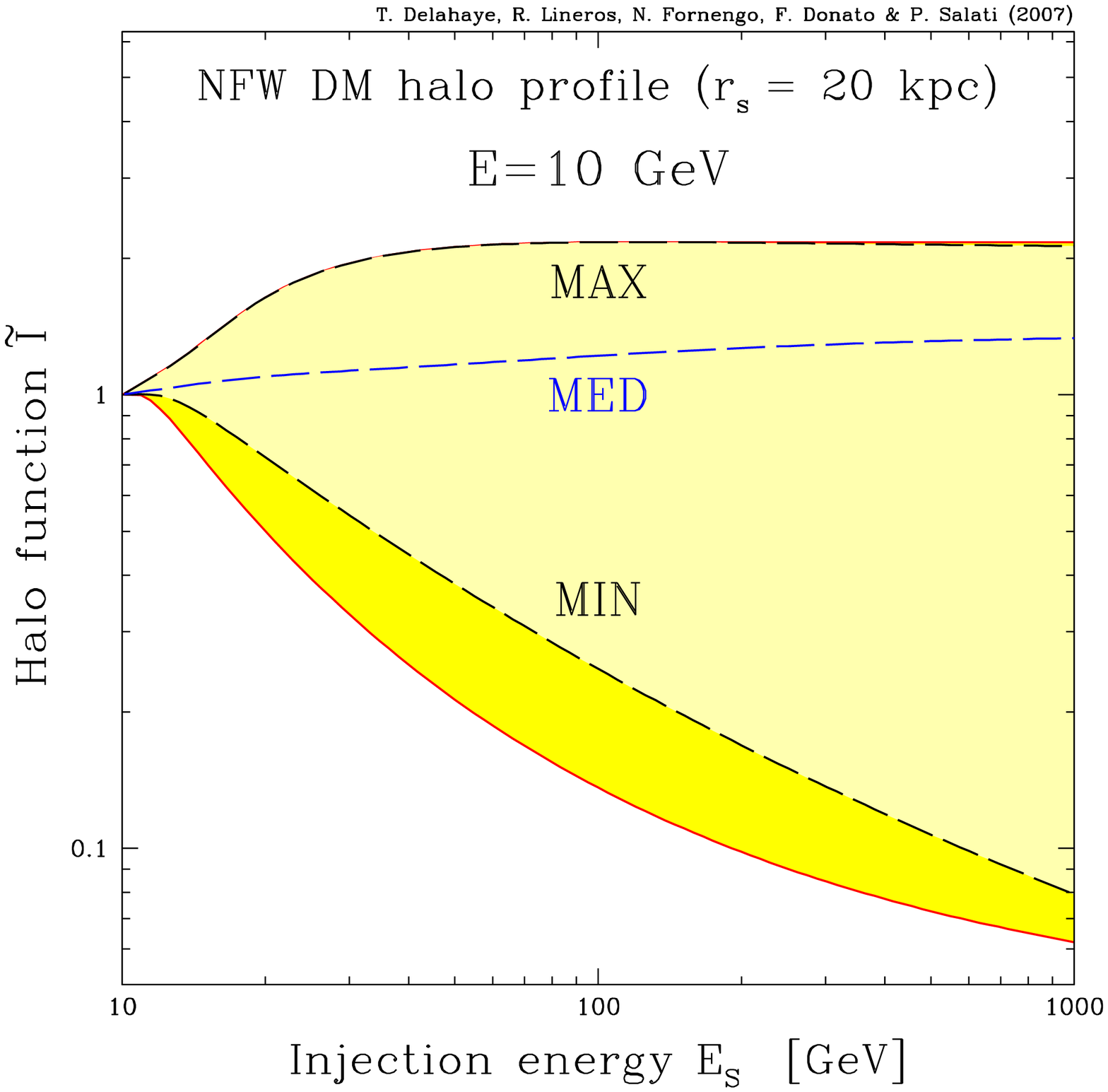}\\
\end{minipage}
%
%
\begin{center}
\noindent
\includegraphics [width=0.5\textwidth]{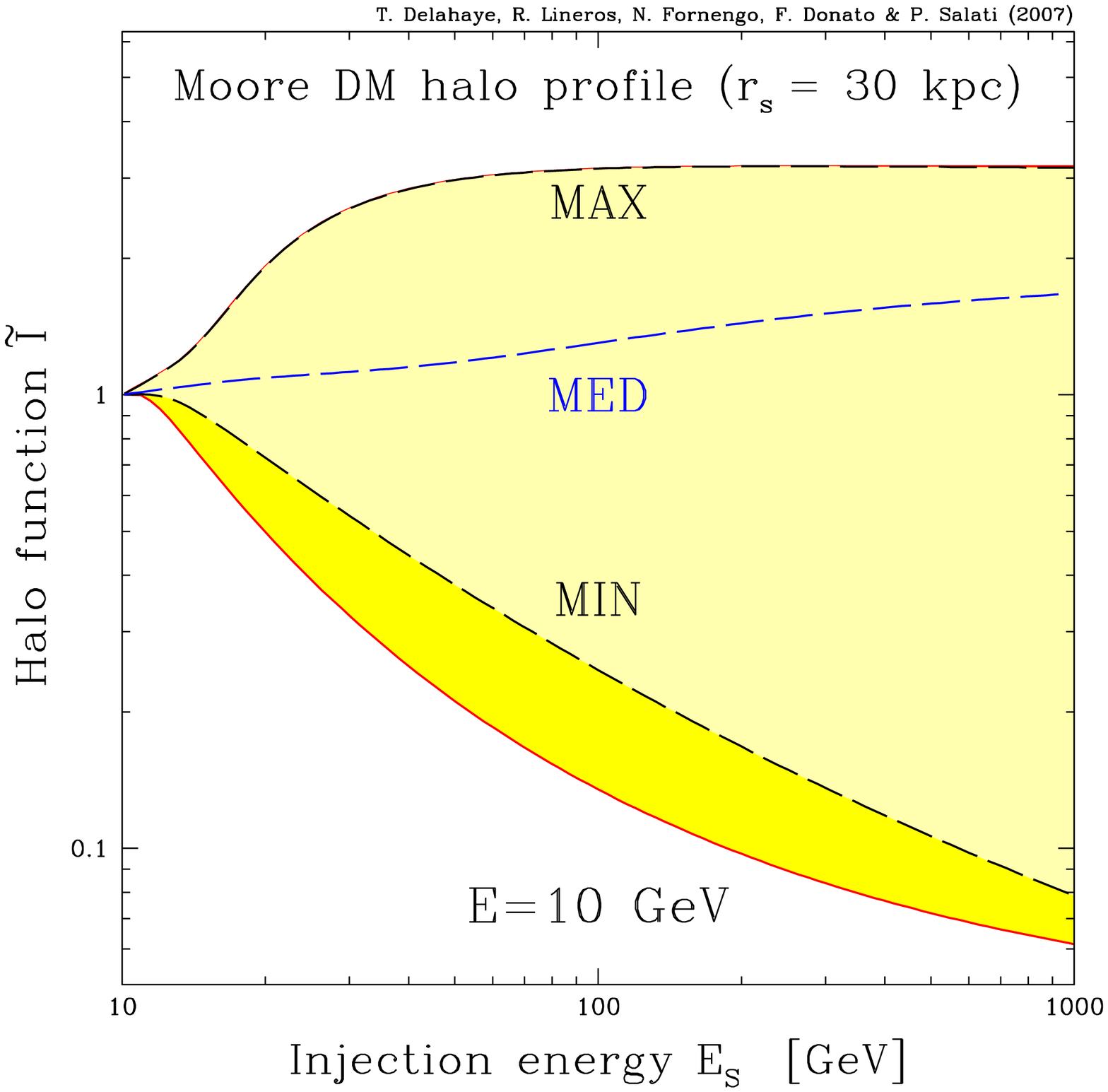}
\end{center}
\vskip -0.50cm \caption{ In each panel, the halo integral
$\tilde{I}$ is plotted as a function of the positron injection
energy $E_{S}$ whereas the energy $E$ at the Earth is fixed at 10
GeV. The galactic DM halo profiles of Tab.~\ref{tab:indices} are
featured.
The curves labeled as MED correspond to the choice of cosmic ray
propagation parameters which best-fit the B/C ratio \cite{parfit}.
The MAX and MIN configurations correspond to the cases which were
identified to produce the maximal and minimal DM antiproton fluxes
\cite{primary_pbar}, while the entire colored band corresponds to
the complete set of propagation models compatible with the B/C
analysis \cite{parfit}. } \label{fig5}
\end{figure*}
%

%
\begin{figure}[t] \centering
\includegraphics[width=\columnwidth]{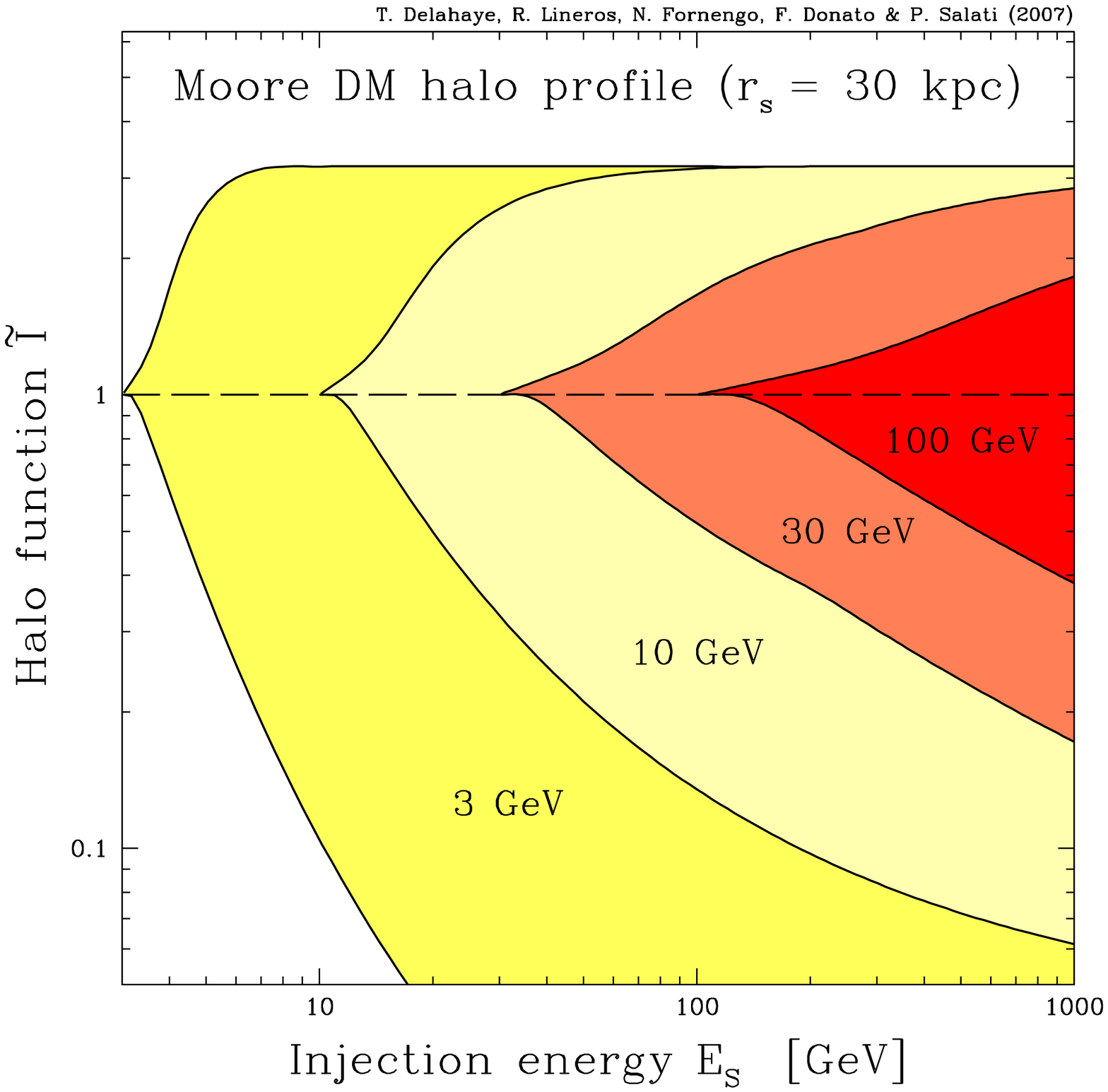}
\caption{Same plot as before where the Moore DM profile has been selected. Four values of the positron
detection energy $E$ have been assumed. The flagella structure of this figure results from
the widening of the uncertainty band as the detection energy $E$ is decreased.
\label{fig6}}
\end{figure}
%

\section{Propagation uncertainties on the halo integral}
\label{sect:3}

Following the prescription which has been given in the previous
section, we can calculate accurately and quickly the halo integral
$\tilde{I}$ using either the Green propagator method or the Bessel
expansion technique according to the typical diffusion length
$\lD$. We are now equipped with a rapid enough method for scanning
the $\sim$ 1,600 different cosmic ray propagation models that have
been found compatible \cite{parfit} with the B/C measurements.
Each model is characterized by the half-thickness $L$ of the
diffusion zone and by the normalization $K_{0}$ and spectral index
$\delta$ of the space diffusion coefficient.  A large variation in
these parameters is found in \cite{parfit} and yet they all lead
to the same B/C ratio. The height $L$ of the diffusion slab lies
in the range from 1 to 15 kpc. Values of the spectral index
$\delta$ extend from 0.46 to 0.85 whereas the ratio ${K_{0}}/{L}$
varies from $10^{-3}$ to $8 \times 10^{-3}$ kpc Myr$^{-1}$.

In this section, we analyze the sensitivity of the positron halo
integral $\tilde{I}$ with respect to galactic propagation. We
would like eventually to gauge the astrophysical uncertainties
which may affect the predictions on the positron DM signal.
A similar investigation -- with only the propagation
configurations that survive the B/C test -- has already been
carried out for secondary \cite{secondary_pbar} and primary
\cite{primary_pbar} antiprotons. In the later case, three specific
sets of parameters have been derived corresponding to minimal,
medium and maximal antiproton fluxes -- see Tab.~\ref{tab:pbar}.

Do these configurations play the same role for positrons~? Can we
single out a few propagation models which could be used later on
to derive the minimal or the maximal positron flux without
performing an entire scan over the parameter space~?
These questions have not been addressed in the pioneering
investigation of \cite{Hooper:2004bq} where the cosmic ray
propagation parameters have indeed been varied but independently
of each other and without any connection to the B/C ratio.

%
\begin{table}[t]
\vskip 0.5cm
{\begin{tabular}{|c||c|c|c|}
\hline
Model  & $\delta$ & $K_0$ [kpc$^2$/Myr] & $L$ [kpc] \\
\hline \hline
MIN  & 0.85 &  0.0016 & 1  \\
MED  & 0.70 &  0.0112 & 4  \\
MAX  & 0.46 &  0.0765 & 15 \\
\hline
\end{tabular}}
\caption{Typical combinations of cosmic ray propagation parameters that are compatible
with the B/C analysis \cite{parfit} and which have been found \cite{primary_pbar} to
correspond respectively to minimal, medium and maximal primary antiproton fluxes.}
\label{tab:pbar}
\end{table}
%
In \citefig{fig5}, we have set the positron detection energy $E$
at a fixed value of 10 GeV and varied the injection energy $E_{S}$
from 10 GeV up to 1 TeV. The three panels correspond to the DM
halo profiles of Tab.~\ref{tab:indices}. For each value of the
injection energy $E_{S}$, we have performed a complete scan over
the 1,600 different configurations mentioned above and have found
the maximal and minimal values of the halo integral $\tilde{I}$
with the corresponding sets of propagation parameters. In each
panel, the resulting uncertainty band corresponds to the yellow
region extending between the two solid red lines. The lighter
yellow domain is demarcated by the long-dashed black curves
labeled MIN and MAX and has a smaller spread. The MED
configuration is featured by the long-dashed blue line.
In \citefig{fig6}, the Moore profile has been chosen  with four
different values of the detection energy $E$. The corresponding
uncertainty bands are coded with different colors and encompass
each other as $E$ increases.

As $E_{S}$ gets close to $E$, we observe that each uncertainty
domain shrinks. In that regime, the diffusion length $\lD$ is very
small and the positron horizon probes only the solar neighborhood
where the DM density is given by $\rho_{\odot}$. Hence the
flagellate structure of \citefig{fig6} and a halo integral
$\tilde{I}$ of order unity whatever the propagation model.

As is clear in \citefig{fig2}, a small half-thickness $L$ of the
diffusion slab combined with a large diffusion length $\lD$
implies a small positron halo integral $\tilde{I}$. The lower
boundaries of the various uncertainty bands in \citefig{fig5} and
\ref{fig6} correspond therefore to parameter sets with $L = 1$
kpc. Large values of $\lD$ are obtained when both the
normalization $K_{0}$ and the spectral index $\delta$ are large --
see \citeeq{definition_lD}. However both conditions cannot be
satisfied together once the B/C constraints are applied. For a
large normalization $K_{0}$, only small values of $\delta$ are
allowed and vice versa. For small values of the detection energy
$E$, the spectral index $\delta$ has little influence on $\lD$ and
the configuration which minimizes the halo integral $\tilde{I}$
corresponds to the large normalization $K_{0} = 5.95 \times
10^{-3}$ kpc$^{2}$ Myr$^{-1}$ and the rather small $\delta =
0.55$. For large values of $E$, the spectral index $\delta$
becomes more important than $K_{0}$ in the control of $\lD$. That
is why in \citefig{fig6}, the lower bound of the red uncertainty
domain corresponds now to the small normalization $K_{0} = 1.65
\times 10^{-3}$ kpc$^{2}$ Myr$^{-1}$ and the large spectral index
$\delta = 0.85$. Notice that this set of parameters is very close
to the MIN configuration of Tab.~\ref{tab:pbar}. For intermediate
values of $E$, the situation becomes more complex. We find in
particular that for $E = 30$ GeV, the halo integral $\tilde{I}$ is
minimal for the former set of parameters as long as $E_{S} \leq
200$ GeV and for the later set as soon as $E_{S} \geq 230$ GeV. In
between, a third propagation model comes into play with the
intermediate values $K_{0} = 2.55 \times 10^{-3}$ kpc$^{2}$
Myr$^{-1}$ and $\delta = 0.75$. It is not possible therefore to
single out one particular combination of $K_{0}$ and $\delta$
which would lead to the minimal value of the halo integral and of
the positron DM signal. The MIN configuration which appeared in
the antiproton analysis has no equivalent for positrons.

The same conclusion holds, even more strongly, in the case of the
upper boundaries of the uncertainty bands.
Whatever the DM halo profile, the panels of \citefig{fig2} feature
a peak in the halo function $\tilde{I}$ for large values of $L$
and for a specific diffusion length $\lDM \sim 7$ kpc. At fixed
$E$ and $E_{S}$, we anticipate that the maximal value for
$\tilde{I}$ will be reached for $L = 15$ kpc and for a diffusion
length $\lD$ as close as possible to the peak value $\lDM$. Two
regimes can be considered at this stage.

\vskip 0.1cm \noindent {\bf (i)} To commence, the diffusion length
$\lD$ is below the critical value $\lDM$ whenever the difference
$v(\epsilon) - v(\epsilon_{S})$ is small enough -- see the
definitions~(\ref{connection_E_pseudo_t}) and
(\ref{definition_lD}). This condition is met in general when $E$
and $E_{S}$ are close to each other or when $E$ is large. The
largest possible value of $\lD$ maximizes $\tilde{I}$ and once
again, we find two propagation models.
For small $E$, the large normalization $K_{0} = 7.65 \times
10^{-2}$ kpc$^{2}$ Myr$^{-1}$ is preferred with $\delta = 0.46$.
We recognize the MAX configuration of Tab.~\ref{tab:pbar} and
understand why the long-dashed black curves labeled MAX in the
panels of \citefig{fig5} are superimposed on the solid red upper
boundaries.
For large $E$, the spectral index $\delta$ dominates the diffusion
length $\lD$ and takes over the normalization $K_{0}$ of the
diffusion coefficient. The best model which maximizes $\tilde{I}$
becomes then $\delta = 0.75$ and $K_{0} = 2.175 \times 10^{-2}$
kpc$^{2}$ Myr$^{-1}$.

\vskip 0.1cm \noindent {\bf (ii)} When the difference $v(\epsilon)
- v(\epsilon_{S})$ is large enough, the diffusion length $\lD$ may
reach the critical value $\lDM$ for at least one propagation model
which therefore maximizes the halo integral.
As $E$ and $E_{S}$ are varied, the peak value of $\tilde{I}$ is
always reached when a scan through the space of parameters is
performed. This peak value corresponds to the maximum of the halo
integral, hence a horizontal upper boundary for each of the
uncertainty bands of \citefig{fig5} and \ref{fig6}.
The set that leads to $\lD = \lDM$ is different for each
combination of $E$ and $E_{S}$ and is not unique. In the case of
the NFW DM profile of \citefig{fig5}, the halo integral
$\tilde{I}$ is maximized by more than 30 models above $E_{S} =
120$ GeV.

The complexity of this analysis confirms that the propagation
configurations selected by B/C do not play the same role for
primary antiprotons and positrons. The two species experience the
propagation phenomena, and in particular energy losses, with
different intensities. As pointed out in Ref.~\cite{Maurin:2002uc},
the average distance traveled by a positron is sensibly lower
than the one experienced by an antiproton produced in the halo.

\section{Positron fluxes}
\label{sect:4}

Now that we have discussed in detail the solution of the
propagation equation, and have identified and quantified the
astrophysical uncertainties on the halo integral $\tilde I$, we
are ready to apply our analysis to the theoretical predictions for
the positron signal at the Earth position. The positron flux is
obtained through Eq.~(\ref{eq:fluxe}). As stated in
Sec.~\ref{sec:source}, we will not adopt specific DM candidates,
but will instead discuss the signals arising from a DM particle
which annihilates into a pure final state. We consider four
different specific DM annihilation channels: direct $e^+ e^-$
production as well as $W^+ W^-$, $b \bar b$ and $\tau^+ \tau^-$.
The DM annihilation cross section is fixed at the value
$2.1~\times~10^{-26}$~cm$^{3}$~s$^{-1}$ and we will consider the
cases of a DM species with mass of 100~GeV and of 500~GeV.
Generic DM candidates, for instance a neutralino or a sneutrino in
supersymmetric models, or the lightest Kaluza--Klein particle in
models with extra--dimensions, will entail annihilation processes
with specific branching ratios into one or more of these benchmark
cases. The positron flux in these more general situations would
simply be a superposition of the results for each specific
annihilation channel, weighted by the relevant branching ratios
and normalized by the actual annihilation cross section.
%
\begin{table}[h!]
\vskip 0.5cm
{\begin{tabular}{|c||c|c|c|}
\hline
Model  & $\delta$ & $K_0$ [kpc$^2$/Myr] & $L$ [kpc] \\
\hline \hline
MED & 0.70 &  0.0112 & 4  \\
M1  & 0.46 &  0.0765 & 15  \\
M2  & 0.55 &  0.00595 & 1 \\
 \hline
\end{tabular}}
\caption{Typical combinations of cosmic ray propagation parameters
that are compatible with the B/C analysis \cite{parfit}. The model
MED has been borrowed from Tab.~\ref{tab:pbar}. Models M1 and M2
respectively maximize and minimize the positron flux over some energy
range -- roughly above 10 GeV -- the precise extent of which depends
on the mass of the DM particle, on the annihilation channel and also
on the DM profile.
Note that M1 is the same as MAX in Tab.~\ref{tab:pbar} but this is
coincidental.}
\label{tab:model}
\end{table}
%

%
\begin{figure*}[t!] \centering
\includegraphics[angle=270, width=0.9\textwidth]{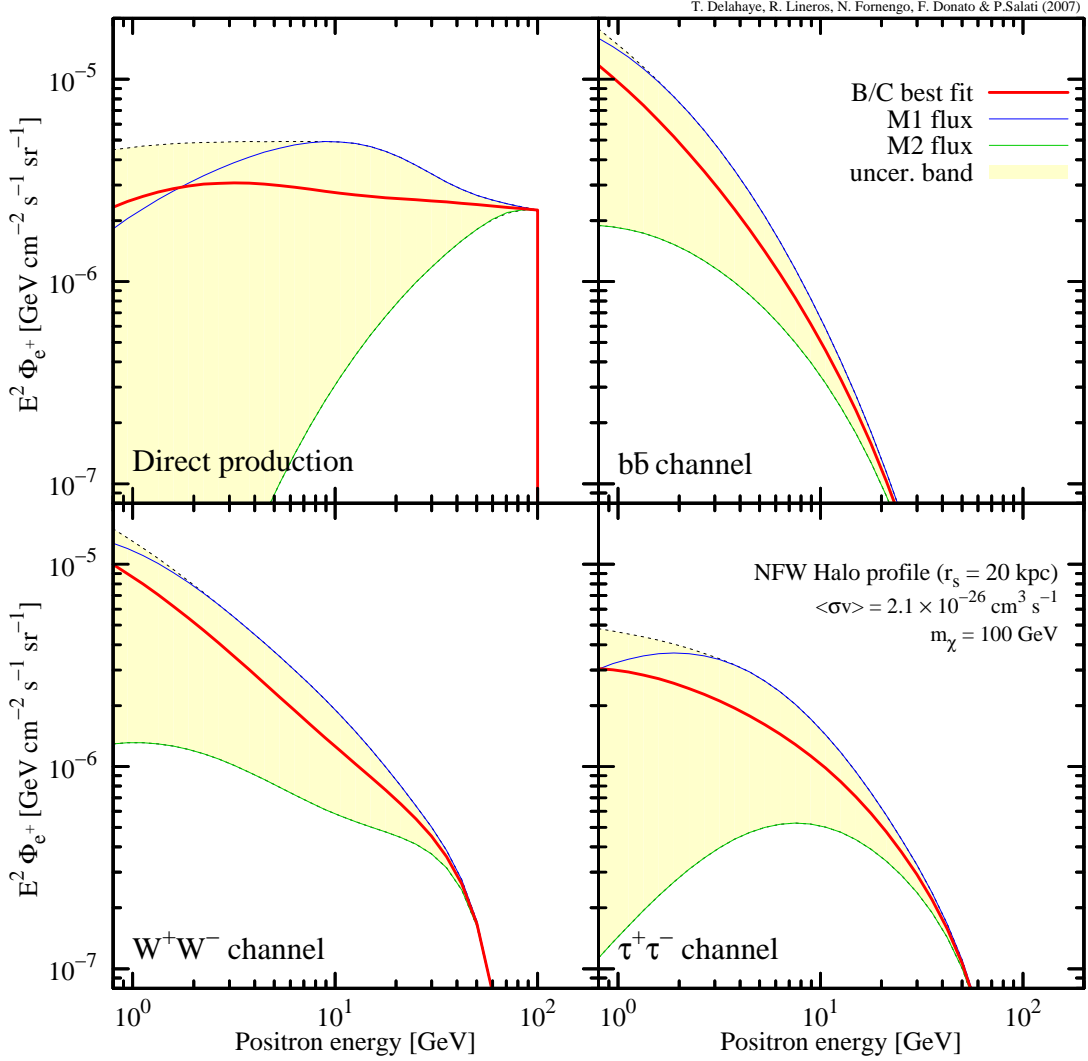}
\caption{
\label{fig:f1-100gev}
Positron flux $E^2 \Phi_{e^+}$ versus the positron energy $E$,
for a DM particle with a mass of 100 GeV and for a NFW profile
-- see Tab.~\ref{tab:indices}.
The four panels refer to different annihilation final states~:
direct $e^+ e^-$ production (top left), $b \bar b$ (top right),
$W^+ W^-$ (bottom left) and $\tau^+ \tau^-$ (bottom right). In
each panel, the thick solid [red] curve refers to the best--fit
choice (MED) of the astrophysical parameters. The upper [blue]
and lower [green] thin solid lines correspond respectively to
the astrophysical configurations which provide here the maximal
(M1) and minimal (M2) flux -- though only for energies above a
few GeV in the case of (M1).
The colored [yellow] area features the total uncertainty band
arising from positron propagation.}
\end{figure*}
%

%
\begin{figure*}[t!]
\centering
\includegraphics[angle=270, width=0.9\textwidth]{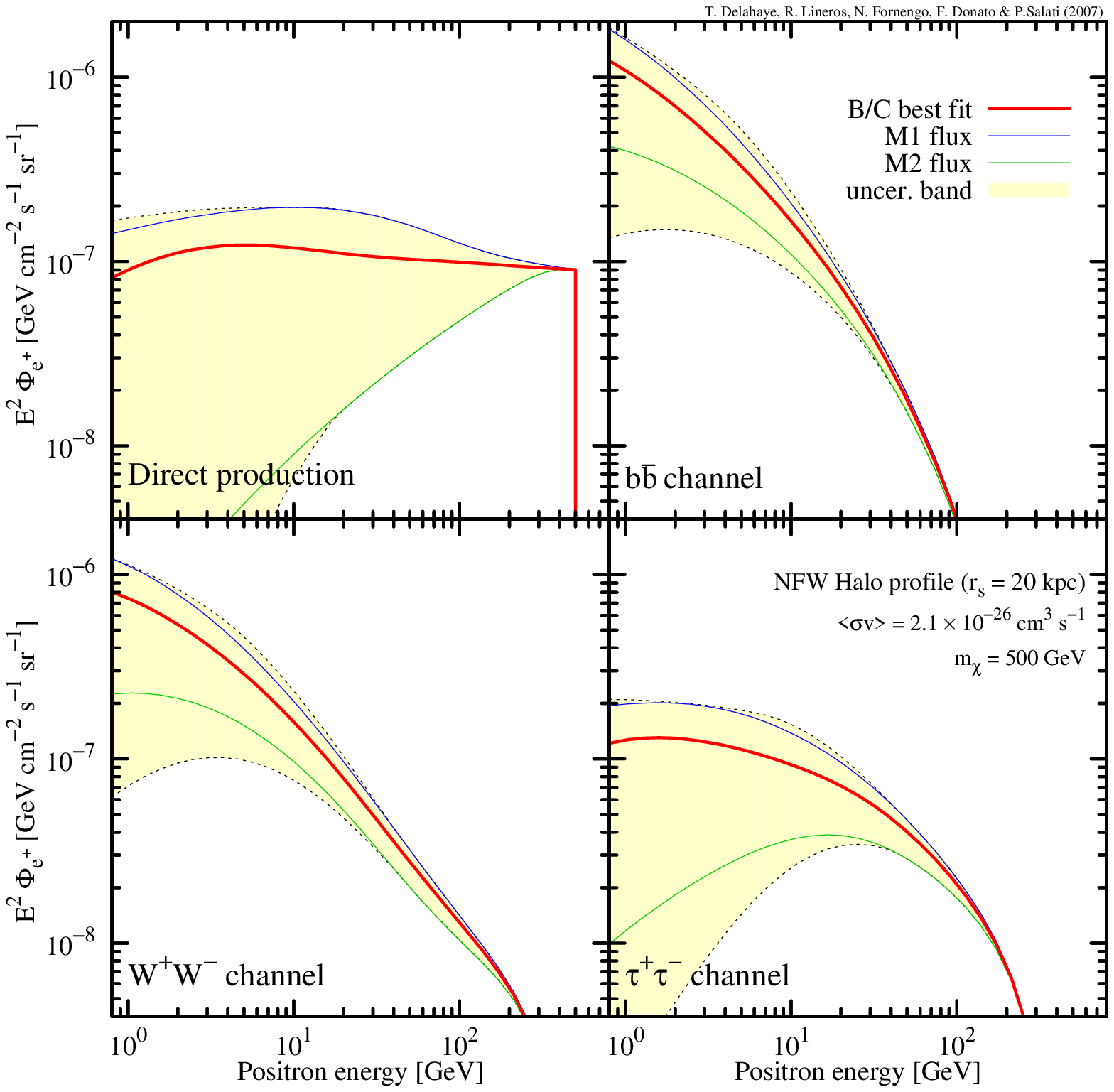}
\caption{
\label{fig:f1-500gev}
Same plot as in Fig.~\ref{fig:f1-100gev} but with a DM particle
mass of $500$ GeV.}
\end{figure*}
%

%
%
In Fig.~\ref{fig:f1-100gev}, the propagated positron
flux~$\fluxe$ -- multiplied by the square of the positron energy~$E$
for convenience -- is featured as a function of $E$ for a 100
GeV DM particle and a NFW density profile.
%
%
The colored [yellow] area corresponds to the total uncertainty band
arising from positron propagation.
In all panels, it enlarges at low positron energy. This may be
understood as a consequence of the behavior of the halo
integral~$\tilde I$ which was analyzed in Sec.~\ref{sect:3}.
Positrons produced at energy $E_{S}$ and detected at energy $E$
originate on average from a sphere whose radius is $\lD$. That
positron sphere enlarges as $E$ decreases and so does
the uncertainty band. As positrons originate further from the
Earth, the details of galactic propagation become more important
in the determination of the positron flux. On the contrary,
high--energy positrons are produced locally and the
halo integral~$\tilde I$ becomes unity whatever the astrophysical
parameters.
Notice also that the uncertainty band can be sizeable and depends
significantly on the positron spectrum at production. In the case
of the $e^{+} e^{-}$ line of the upper left panel, the positron
flux~$\fluxe$ exhibits a strongly increasing uncertainty as
$E$ is decreased from $m_{\chi}$ down to 1~GeV. That uncertainty
is one order of magnitude at $E = 10$~GeV, and becomes larger than
2 orders of magnitude below 1~GeV.
Once again, the positron sphere argument may be invoked. At fixed
detected energy $E$, the radius $\lD$ increases with the injected
energy $E_{S}$. We therefore anticipate a wider uncertainty band as
the source spectrum gets harder. This trend is clearly present in
the panels of Fig.~\ref{fig:f1-100gev}. Actually direct production
is affected by the largest uncertainty, followed by the
$\tau^+ \tau^-$ and $W^+ W^-$ channels where a positron is
produced either directly from the $W^+$ or from the leptonic
decays.
In the $b \bar b$ case, which is here representative of all quark
channels, a softer spectrum is produced since positrons arise mostly
from the decays of charged pions originating from the quark
hadronization. Most of the positrons have already a low energy
$E_{S}$ at injection and since they are detected at an energy
$E \sim E_{S}$, they tend to have been produced not too far from
he Earth, hence a lesser dependence to the propagation uncertainties.
The astrophysical configuration M2 -- see Tab.~\ref{tab:model} --
provides the minimal positron flux. It corresponds to the lower
boundaries of the yellow uncertainty bands of Fig.~\ref{fig:f1-100gev}.
The M1 configuration maximizes the flux at high energies. For direct
production and to a lesser extent for the $\tau^+ \tau^-$ channel,
that configuration does not reproduce the upper envelope of the
uncertainty band in the low energy tail of the flux. As discussed
in Sec.~\ref{sect:3}, the response of $\fluxe$ to the propagation
parameters depends on the detected energy $E$ in such a way that
the maximal value cannot be reached for a single astrophysical
configuration.
Finally, taking as a reference the median flux, the uncertainty bands
extend more towards small values of the flux. In all channels, the
maximal flux is typically a factor of $\sim$~1.5--2 times larger
than the median prediction. The minimal flux features larger deviations
with a factor of 5 for the $b \bar b$~channel at $E = 1$~GeV, of 10
for $W^+ W^-$ and of 30 for $\tau^+ \tau^-$.

%
%
Fig.~\ref{fig:f1-500gev} is similar to Fig.~\ref{fig:f1-100gev}
but with a heavier DM species of 500 GeV instead of 100 GeV.
Since the mass $m_{\chi}$ is larger, so is on average the
injected energy $E_{S}$. Notice that at fixed positron energy $E$
at the Earth, the radius $\lD$ of the positron sphere increases with
$E_{S}$. We therefore anticipate that the propagated fluxes are affected
by larger uncertainties for heavy DM particles. Again, the maximal flux
does not exceed twice the median flux, while the minimal configurations
are significantly depressed. At the reference energy $E = 1$~GeV,
reductions by a factor of 10 between the median and minimal predictions
are obtained for the $b \bar b$ channel and amount to a factor of 20
in the $W^+ W^-$ case. They reach up to 2 orders of magnitude for
the direct positron production.
In this large DM mass regime, the astrophysical configuration M2
does not reproduce by far the lower bound of the uncertainty band
as it did for the 100~GeV case. The message is therefore twofold.
\begin{itemize}
\item[(i)] Once the positron spectrum at the source is chosen -- and
the corresponding branching ratios have been defined -- the correct
determination of the uncertainty which affects the flux at the Earth
requires a full scan of the propagation parameter space for each energy
$E$. The use of representative astrophysical configurations such as
M1 and M2 would not provide the correct uncertainty over the entire
range of positron energy $E$.
\item[(ii)] However, specific predictions have to be performed for
a given model of DM particle and a fixed set of astrophysical parameters.
This is why fits to the experimental data should be performed for each
propagation configuration over the entire range of the measured positron
energies $E$. The best fit should correspond to a unique set of
astrophysical parameters. This procedure is the only way to reproduce
properly the correct and specific spectral shape of the flux.
\end{itemize}

%
\begin{figure*}[t!] \centering
\includegraphics[angle=270, width=0.9\textwidth]{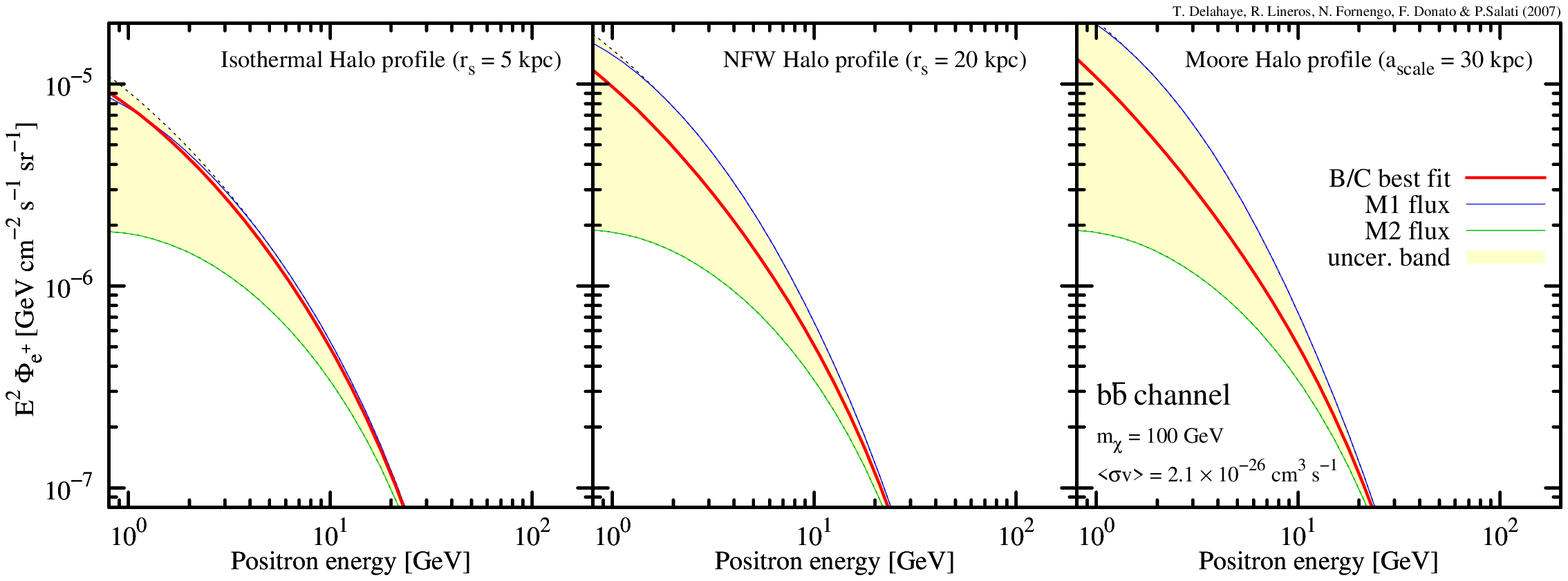}
\includegraphics[angle=270, width=0.9\textwidth]{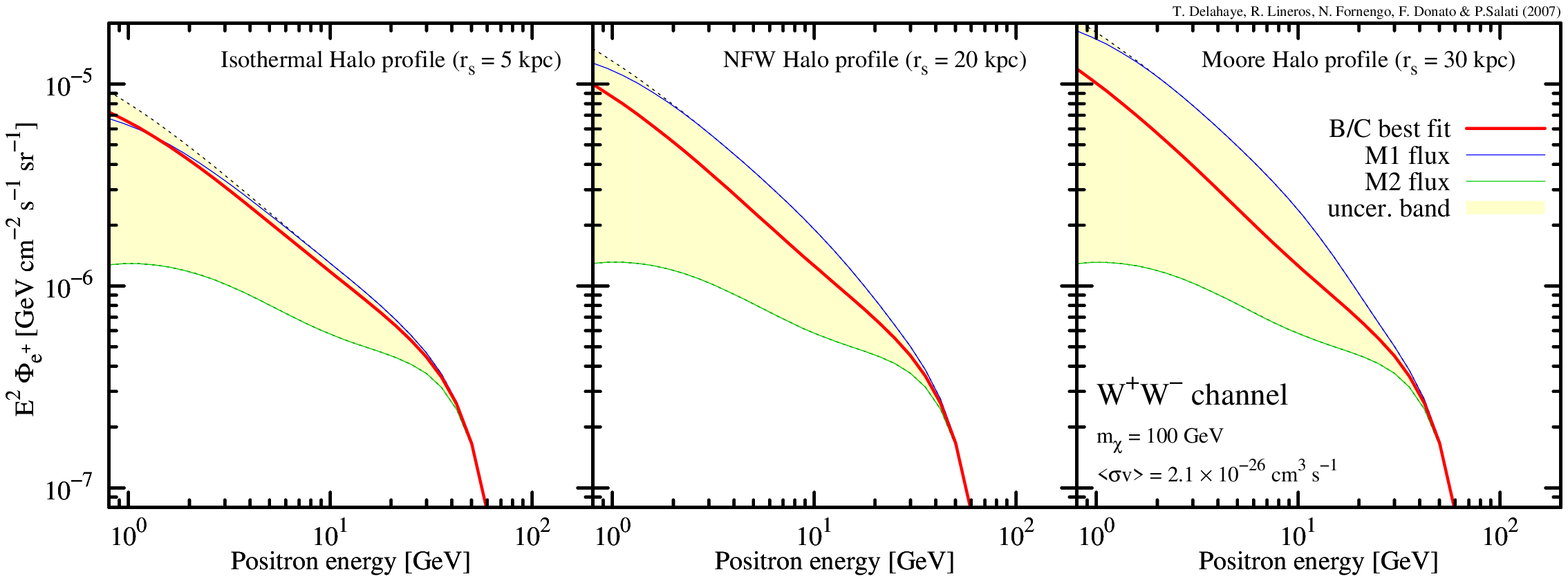}
\caption{
\label{fig:f2-bb}
Positron flux $E^2 \Phi_{e^+}$ versus the positron energy $E$, for a DM particle
mass of 100~GeV and for different halo density profiles~: cored isothermal
sphere~\cite{bahcall} (left panels), NFW~\cite{nfw} (central panels) and
Moore~\cite{moore} (right panels) -- see Tab.~\ref{tab:indices}.
The upper and lower rows correspond respectively to a $b \bar{b}$ and
$W^{+} W^{-}$ annihilation channel. In each panel, the thick solid [red]
curve refers to the best--fit choice (MED) of the astrophysical parameters.
The upper [blue] and lower [green] thin solid lines stand for the astrophysical
configurations M1 and M2 of Tab.~\ref{tab:model}.
The colored [yellow] area indicates the total uncertainty band arising from
positron propagation.}
\end{figure*}
%

%
\begin{figure*}[t] \centering
\includegraphics[angle=270, width=0.9\textwidth]{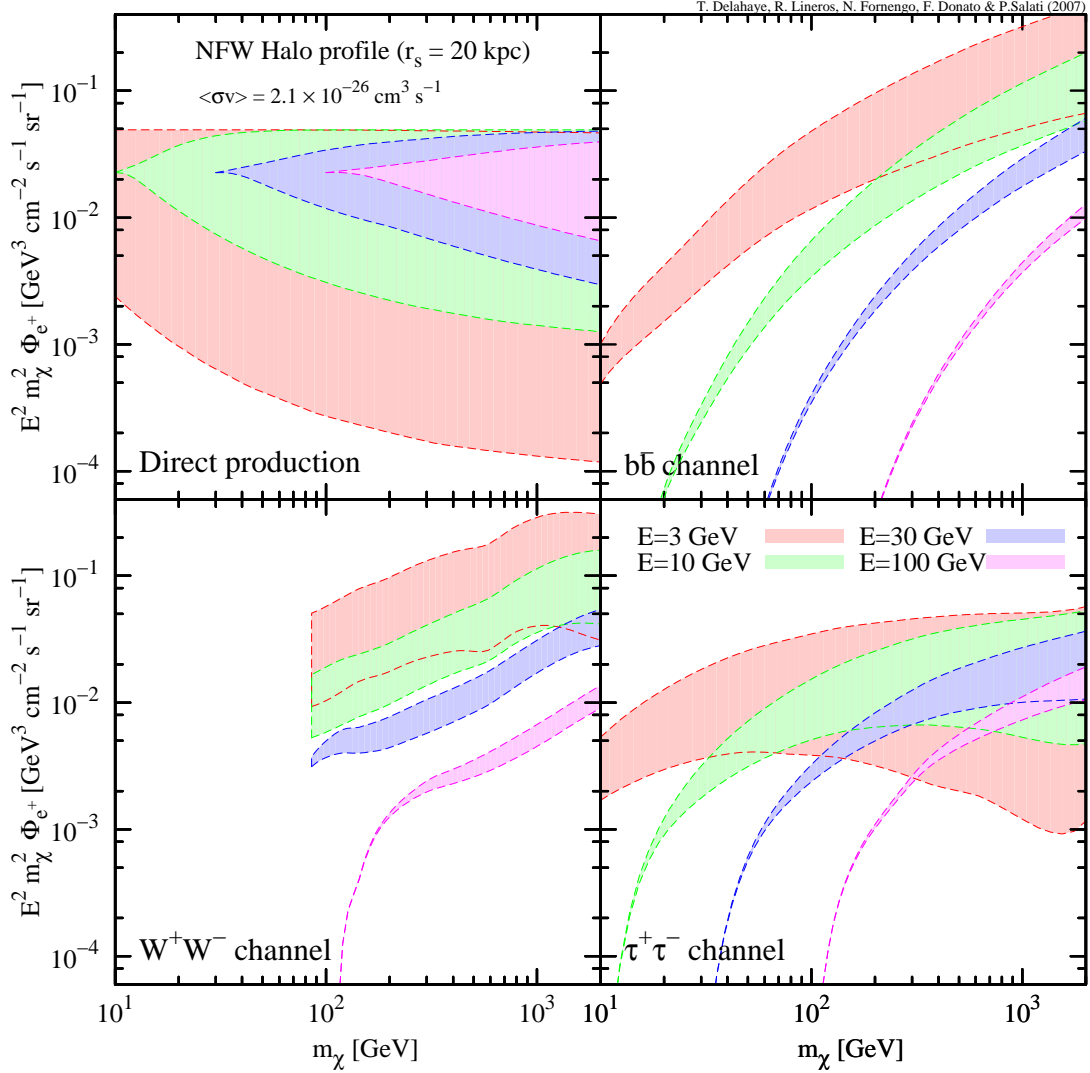}
\caption{
\label{fig:f5-fluxmass}
For fixed values of the detected energy $E$, the uncertainty bands
on the positron flux $E^2 m_\chi^2 \Phi_{e^+}$ are shown as a function
of the mass $m_\chi$ of the DM particle. The energies considered in the
figure are $E = 3$, 10, 30 and 100 GeV. Each band refers to one of those
values and starts at $m_{\chi} = E$.}
\end{figure*}
%

%
\begin{figure*}[t] \centering
\includegraphics[angle=270, width=0.9\textwidth]{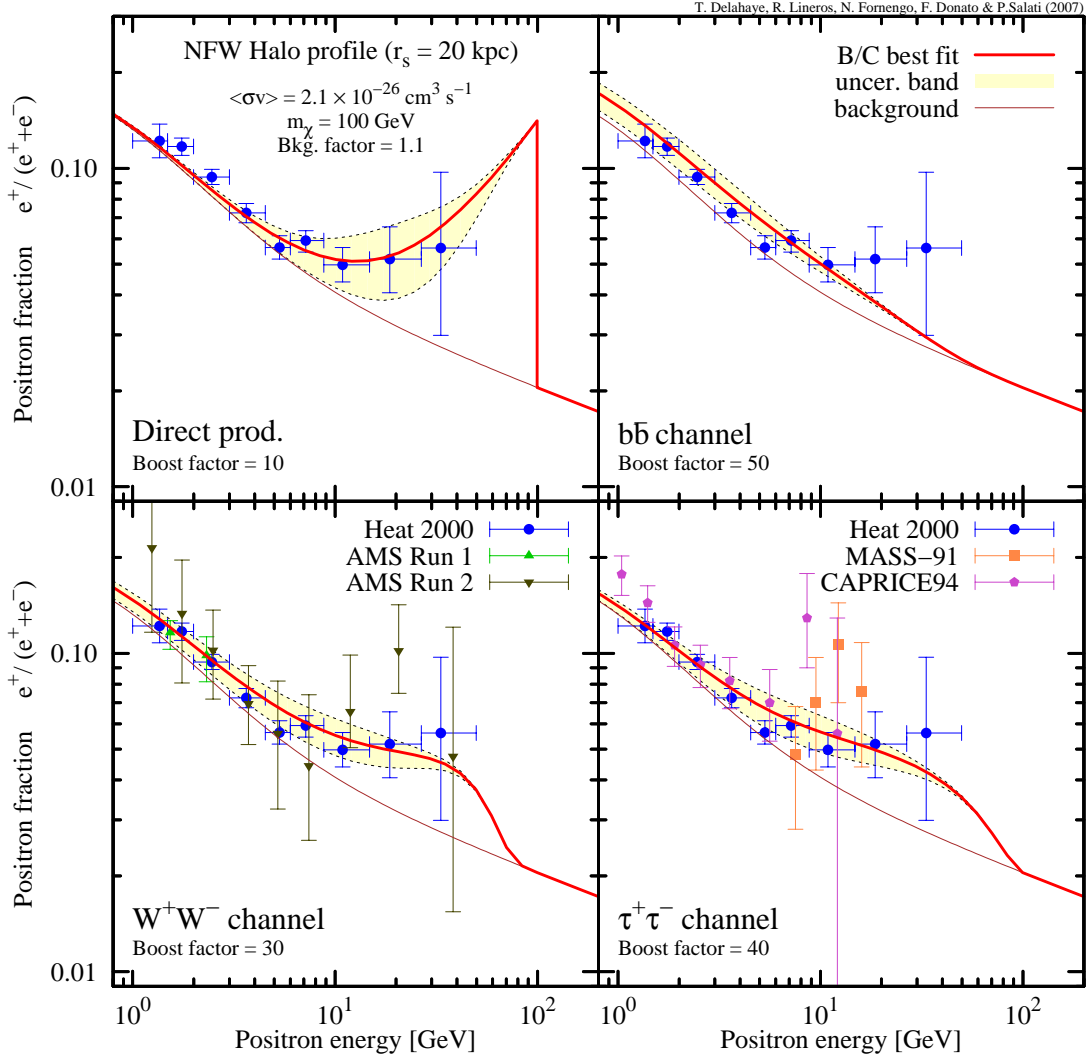}
\caption{
\label{fig:f3-heat-pf-100gev}
Positron fraction $e^+/(e^- + e^+)$ versus the positron detection energy $E$.
Notations are as in Fig.~\ref{fig:f1-100gev}. In each panel, the thin [brown]
solid line stands for the background \cite{baltz_edsjo99, strongmoska} whereas
the thick solid [red] curve refers to the total positron flux where the signal
is calculated with the best--fit choice (MED) of the astrophysical parameters.
Experimental data from HEAT~\cite{heat}, AMS~\cite{ams01,ams02},
CAPRICE~\cite{2000ApJ...532..653B} and MASS~\cite{2002A&A...392..287G} are also plotted.}
\end{figure*}
%

%
\begin{figure*}[t] \centering
\includegraphics[angle=270, width=0.9\textwidth]{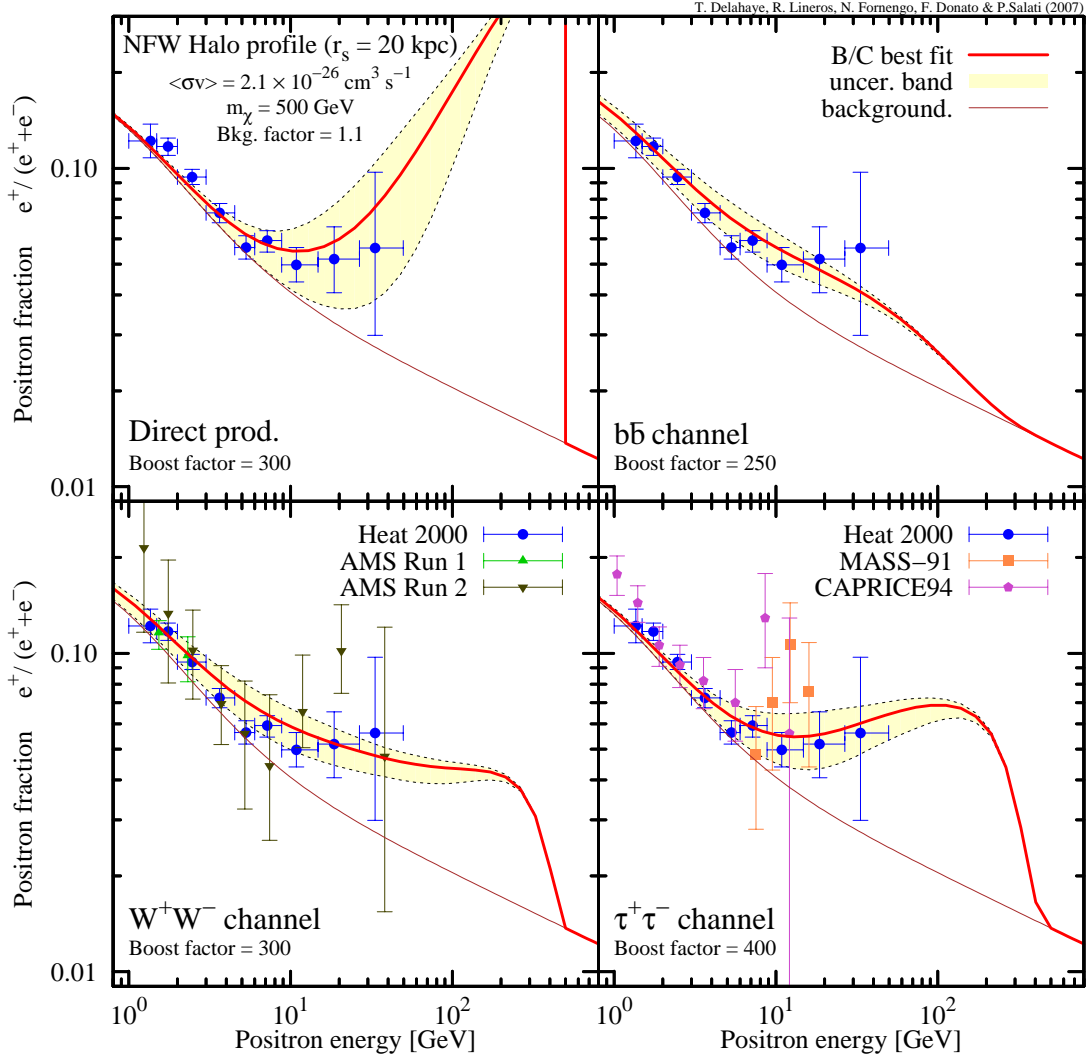}
\caption{
\label{fig:f3-heat-pf-500gev}
Same plot as in Fig.~\ref{fig:f3-heat-pf-100gev} but with a mass of the DM
particle of $500$ GeV.}
\end{figure*}
%

%
\begin{figure*}[t] \centering
\includegraphics[angle=270, width=0.9\textwidth]{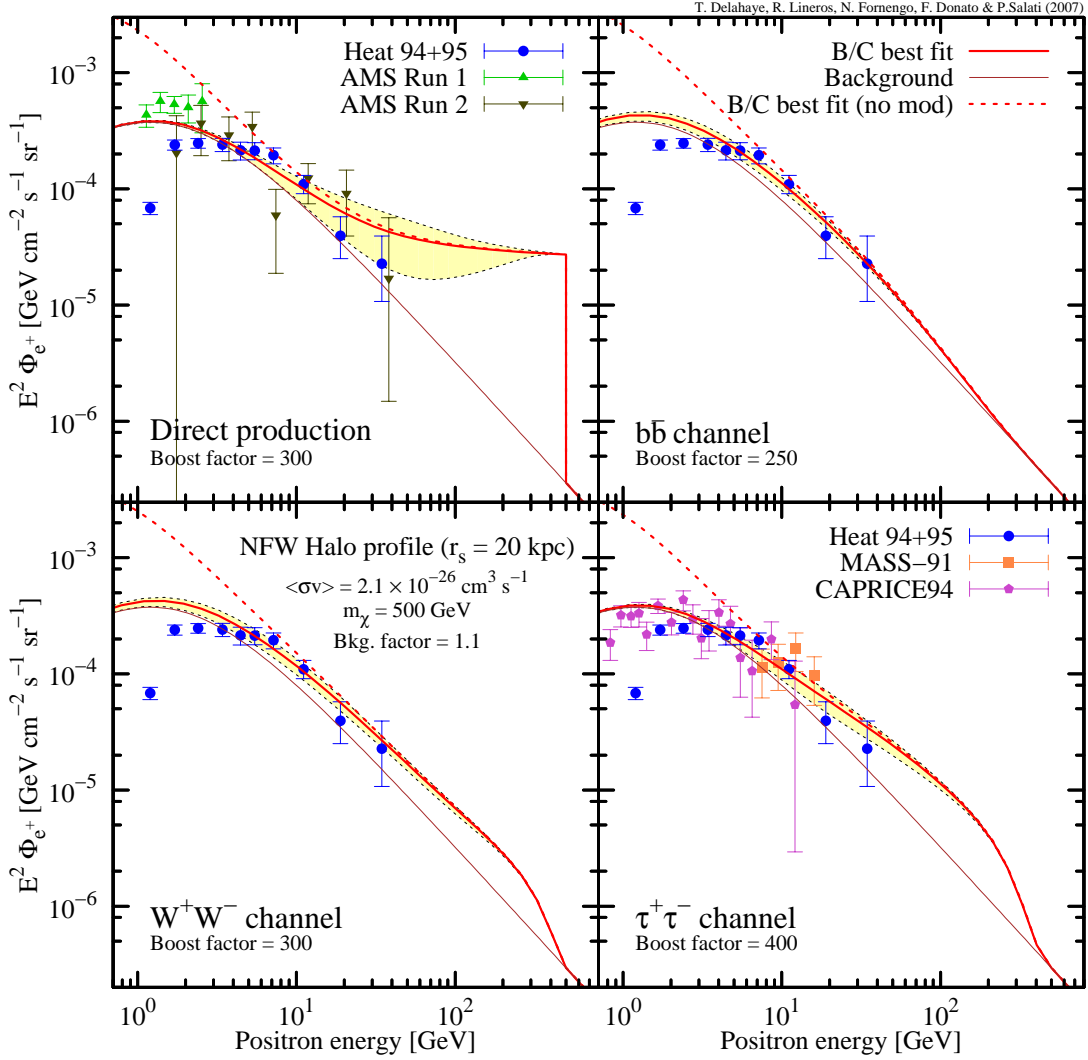}
\caption{
\label{fig:f3-heat-flux-500gev}
Positron flux $E^2 \Phi_{e^+}$ (not fraction) versus the positron energy $E$, for a
500 GeV DM particle. Notations are the same as in Fig.~\ref{fig:f3-heat-pf-100gev}.
Experimental data from HEAT~\cite{heat}, AMS~\cite{ams01,ams02}, CAPRICE~\cite{2000ApJ...532..653B}
and MASS~\cite{2002A&A...392..287G} are plotted.}
\end{figure*}
%

%
\begin{figure*}[t] \centering
\includegraphics[angle=270, width=0.9\textwidth]{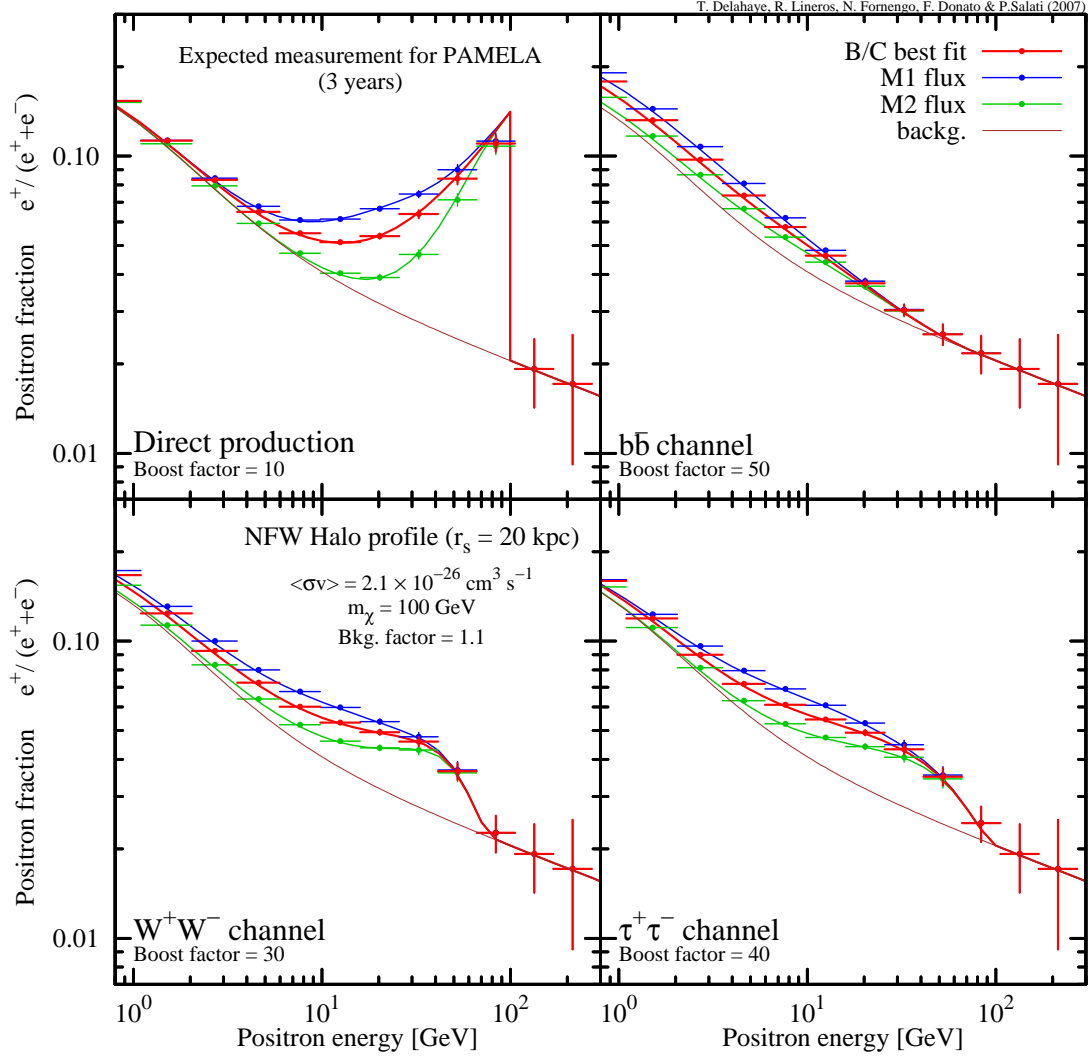}
\caption{
\label{fig:f4-pam-100gev}
Predictions for PAMELA for a 3--year mission.
The positron fraction $e^+/(e^- + e^+)$ and its statistical uncertainty are plotted
against the positron energy $E$ for a 100 GeV DM particle and a NFW profile.
Notations are the same as in Fig.~\ref{fig:f3-heat-pf-100gev}.
The thick solid curves refer respectively to the total positron flux calculated
with the M1 (upper [blue]), MED (median [red]) and M2 (lower [green]) sets
of propagation parameters.}
\end{figure*}
%

%
\begin{figure*}[t] \centering
\includegraphics[angle=270, width=0.9\textwidth]{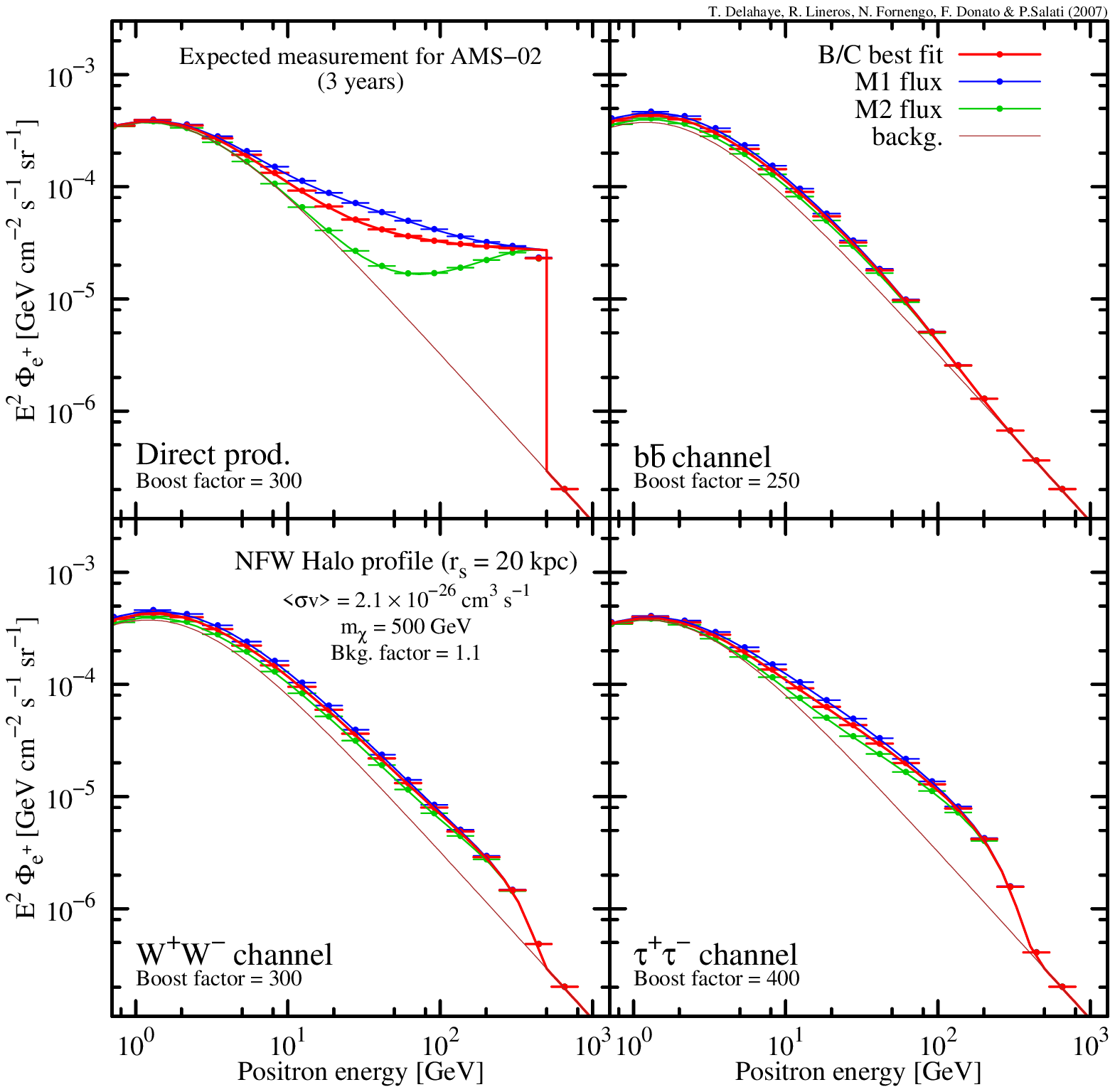}
\caption{
\label{fig:f7-ams-flux-500gev}
Predictions for AMS-02 for a 3--year mission.
The positron flux $E^2 \Phi_{e^+}$ and its statistical uncertainty are featured as
a function of the positron energy $E$ for a 500 GeV DM species and a NFW profile.
Notations are the same as in Fig.~\ref{fig:f3-heat-flux-500gev}.
The thick solid curves refer respectively to the total positron flux calculated
with the M1 (upper [blue]), MED (median [red]) and M2 (lower [green]) sets
of propagation parameters.}
\end{figure*}
%

%
%
The effect induced by different DM profiles is presented in
Fig.~\ref{fig:f2-bb}, where the positron fluxes for the $b\bar b$
and $W^+ W^-$ channels are reproduced for the DM distributions of
Tab.~\ref{tab:indices}. The mass of the DM particle is fixed at
$m_\chi~=~100$~GeV.
Notice how steeper profiles entail larger uncertainties, especially
for the upper bound. This is mostly due to the fact that for large
values of $L$ -- for which larger fluxes are obtained -- the positron
flux is more sensitive to the central region of the Galaxy, where singular
profiles like the NFW and Moore distributions have larger densities and
therefore induce larger annihilation rates.
On the contrary, the lower envelope of the uncertainty band is not
affected by the variation of the halo profile. In this case, with
typically small heights~$L$, positrons reach the solar system from
closer regions, where the three halo distributions are very
similar and do not allow to probe the central part of the Milky Way.

%
%
Fig.~\ref{fig:f5-fluxmass} depicts the information on the positron
flux uncertainty from a different perspective. The flux~$\fluxe$
and its uncertainty band are now featured for fixed values of the
detected energy~$E$ whereas the DM particle mass is now varied.
The flux~$\fluxe$ is actually rescaled by the product
$E^2~m_\chi^2~\fluxe$ for visual convenience. Each band corresponds
to a specific detected energy $E$ and consequently starts at
$m_\chi = E$.
In the case of the $W^+ W^-$ channel, the bands start at
$m_\chi~=~m_W$ because this channel is closed for DM masses below
that threshold.
The behavior of these bands can be understood from Fig.~\ref{fig6},
where the halo function~$\tilde I$ is plotted for the same detected
energies, as a function of the injection energy~$E_S$. In the case
of direct positron production, there is a simple link between the two
figures, because the source spectrum in this case is just a line
at $E_S~=~m_\chi$. For the other channels the situation is more
involved since we have a continuous injection spectrum with
specific features as discussed above.
The main information which can be withdrawn from
Fig.~\ref{fig:f5-fluxmass} is that at fixed detection energy,
the larger the DM mass, the larger the uncertainty. Let us take
for instance a detection energy of $E = 3$~GeV. For direct
production, where $E_S~=~m_\chi$, increasing the DM mass translates
into a larger radius $\lD$ of the positron sphere. As a consequence,
the uncertainty band enlarges for increasing masses. This occurs
for all the annihilation channels, but is less pronounced for soft
spectra as in the $b \bar b$ case. Similar conclusions hold for all
the other values of $E$.

%
%
Comparison with available data is presented in
Fig.~\ref{fig:f3-heat-pf-100gev}, \ref{fig:f3-heat-pf-500gev} and
\ref{fig:f3-heat-flux-500gev}.
%
%
In Fig.~\ref{fig:f3-heat-pf-100gev}, the positron fraction \beq
{\displaystyle \frac{e^+}{e^+ \, + \, e^-}} \equiv {\displaystyle
\frac{\fluxe^{\rm TOT}} {\Phi_{e^-} \, + \, \fluxe^{\rm TOT}}}
\label{fraction} \eeq is plotted as a function of the positron
energy $E$. The total positron flux $\fluxe^{\rm TOT}$ at the
Earth encompasses the annihilation signal and a background
component for which we use the results of Ref.~\cite{strongmoska}
as parameterized in Ref.~\cite{baltz_edsjo99}) -- see the thin
solid [brown] lines. The electron flux is denoted by $\Phi_{e^-}$.
The mass of the DM particle is 100~GeV and a NFW profile has been
assumed. The data from HEAT~\cite{heat}, AMS~\cite{ams01,ams02},
CAPRICE~\cite{2000ApJ...532..653B} and
MASS~\cite{2002A&A...392..287G} are indications of a possible
excess of the positron fraction for energies above 10~GeV. Those
measurements may be compared to the thick solid [red] line that
corresponds to the MED configuration. In order to get a reasonable
agreement between our results and the observations, the
annihilation signal has been boosted by an energy--independent
factor ranging from 10 to 50 as indicated in each panel.
%
%
At the same time, the positron background -- for which we do not
have an error estimate yet -- has been shifted upwards from its
reference value of Ref.~\cite{baltz_edsjo99} by a small amount of 10\%.
%
%
As is clear in the upper left panel, the case of direct production
offers a very good agreement with the potential HEAT excess. Notice
how well all the data points lie within the uncertainty band.
A boost factor of 10 is enough to obtain an excellent agreement
between the measurements and the median flux. A smaller value would
be required for a flux at the upper envelope of the uncertainty
band. The $W^+ W^-$ and $\tau^+ \tau^-$ channels may also reproduce
reasonably well the observations, especially once the uncertainty is
taken into account, but they need larger boost factors of the order
of 30 to 40. On the contrary, softer production channels, like
the $b \bar b$ case, are unable to match the features of the
putative HEAT excess for this value of the DM particle mass.
%
%
For all annihilation channels, the uncertainty bands get thinner at
high energies for reasons explained above. They surprisingly tend to
shrink also at low energies, a regime where the positron horizon is
the furthest and where the details of galactic propagation are expected
to be the most important. Actually, the annihilation signal turns
out to be completely swamped in the positron background. In particular,
the signal from direct production stands up over the background only
for energies larger than 5~GeV. The corresponding uncertainty on the
positron fraction is at most of the order of 50\% for energies between
10 and 20~GeV. In the other cases, the uncertainty bands are even thinner.
%
%
Beware finally of the positron background which should also be affected
by uncertainties due to secondary production processes and propagation.
These uncertainties are not currently available and there is clearly
a need to estimate them in order to properly shape theoretical
predictions and to perform better study of the current and forthcoming
data. Such an investigation would involve a comprehensive analysis
and is out of the scope of the present article.

%
%
Somehow different is the situation for larger masses of the DM
candidate. Fig.~\ref{fig:f3-heat-pf-500gev} features the same
information as Fig.~\ref{fig:f3-heat-pf-100gev}, but now for
$m_\chi=500$~GeV. In this case, all the annihilation channels
manage to reproduce the experimental data, even the softest one
$b \bar b$. For direct production, the positron fraction is very
large at energies above 40~GeV, where no data are currently
available. This feature would be a very clear signature of DM
annihilating directly into $e^+ e^-$ pairs, with strong
implications also on the nature of the DM candidate. For instance,
bosonic dark matter would be strongly preferred, since Majorana
fermionic DM, like the neutralino, possesses a very depressed
cross section into light fermions because of helicity suppression
in the non--relativistic regime.
Astrophysical uncertainties on the signal in this case show up
more clearly than for the case of a lighter DM species, but still
they are not very large. The drawback of having a heavier relic
is that now the boost factors required to match the data are
quite large. In Fig.~\ref{fig:f3-heat-pf-500gev} they range from
250 for the soft channel to 400 for the $\tau^+ \tau^-$ case. Such
large boost factors appear to be disfavored, on the basis of the
recent analysis of Refs.~\cite{Lavalle:2006vb,Lavalle:1900wn}.

%
%
In Fig.~\ref{fig:f3-heat-flux-500gev}, the positron flux
(not the fraction) is compared to the available experimental
data for a 500 GeV DM particle and a NFW profile.
The solid thin [brown] line features the positron background which
we shifted upwards by 10\% with respect to the reference value of
Ref.~\cite{baltz_edsjo99}.
The thick solid [red] line encompasses both that background and
the annihilation signal which we calculated with the best--fit
choice (MED) of the astrophysical parameters. Both curves have
been derived assuming solar modulation implemented through the
force field approximation with a Fisk potential $\phi_{\rm F}$
of 500 MV.
The dashed [red] line instead corresponds to the total positron
interstellar flux without solar modulation. Notice that this curve
is superimposed on the thick [red] line above $\sim$ 10 GeV, a regime
where cosmic ray propagation is no longer affected by the solar wind.
A reasonably good agreement between the theoretical predictions and
the data is obtained, especially once the theoretical uncertainties
on the annihilation signal are taken into account. Notice that the
spread of each uncertainty band is fairly limited as we already
pointed out for the positron fraction. The reasons are the same.

Prospects for the future missions are shown in
Fig.~\ref{fig:f4-pam-100gev} and \ref{fig:f7-ams-flux-500gev}.
%
%
In Fig.~\ref{fig:f4-pam-100gev}, a 100 GeV DM particle and a
NFW halo profile have been assumed.
The median [red] curve corresponds to the prediction for the
best--fit MED choice of astrophysical parameters whereas the upper
[blue] and lower [green] lines correspond respectively to the M1
and M2 propagation models -- see Tab.~\ref{tab:model}.
Since we are dealing with predictions which will eventually be
compared to the measurements performed over an entire range of
positron energies, we have to choose specific sets of propagation
parameters as discussed above in this Section.
The upper and lower curves therefore do not represent the maximal
uncertainty at each energy -- though they may do so in some limited
energy range -- but instead they are ``true'' predictions for a
specific set of propagation parameters.
Fig.~\ref{fig:f4-pam-100gev} summarizes our estimate of
the capabilities of the PAMELA detector~\cite{Boezio:2004jx} after
3 years of running. We only plotted statistical errors. We reach
the remarkable conclusion that not only will PAMELA have the
capability to disentangle the signal from the background, but also
to distinguish among different astrophysical models, especially
for hard spectra.
Our conclusion still holds for the $b \bar b$ soft spectrum
for which the M1, MED and M2 curves of the upper right panel
differ one from each other by more than a few standard deviations.
PAMELA could be able to select among them, even when systematical
errors are included.

%
%
In Fig.~\ref{fig:f7-ams-flux-500gev}, the case of a 500~GeV DM particle
is confronted with the sensitivity of AMS-02 for a 3--year flight.
The possibility to disentangle the signal from the background is also
clearly manifest here, even once the astrophysical uncertainties are
included -- provided though that boost factors of the order of 200 to
400 are possible.
But, unless direct production is the dominant channel, a clear distinction
among the various astrophysical models will be very difficult because
the M1 and M2 configurations are closer to the MED curve now than in the previous
case of a lighter DM species. Comparison between Figs.~\ref{fig:f4-pam-100gev}
and \ref{fig:f7-ams-flux-500gev} clearly exhibits that at least below
the TeV scale, the effect of the mass $m_{\chi}$ should not limit the
capability of disentangling the annihilation signal from the background.
More problematic is our potential to distinguish among different
astrophysical models when the DM mass sizeably exceeds the 100~GeV
scale.

\section{Conclusions}
\label{sect:conclusions}

We have analyzed the positron signal from DM annihilation in the galactic halo,
focusing our attention to the determination of the astrophysical  uncertainties on the positron
flux due to the positron propagation inside the galactic medium.

Propagation of galactic cosmic rays has been treated in a two--zone
model \cite{parfit} and we have solved the diffusion
equation for primary positrons both in the Green function
formalism and with the Bessel expansion method. We find that the
most efficient way of dealing with positron propagation is to
adopt the Green function method for values of the diffusion length
$\lD = \sqrt{4 K_{0} \tilde{\tau}}$ smaller than $\sim 3$ kpc, and
to employ the Bessel function technique whenever $\lD $ becomes larger.
In this way the radial boundaries of the diffusion region (which are
neglected in the Green function approach) can be properly coped
with by the Bessel expansion method.

The propagation uncertainties on the halo integral have been calculated for the  $\sim$ 1,600
different cosmic ray propagation models that have been found compatible \cite{parfit} with the
B/C measurements. These uncertainties are strongly dependent on the source and detection
energies, $E_S$ and $E$.  As $E_{S}$ gets close to $E$, we observe that each uncertainty domain
shrinks. In that regime, the diffusion length $\lD$ is very small and the positron horizon probes
only the solar neighborhood. In the opposite case, the uncertainty can be as large as one order of magnitude
or even more.  As positrons originate further from the Earth, the details of galactic propagation
become more important in the determination of the positron flux. On the contrary, high--energy
positrons are produced locally and the halo integral~$\tilde I$ becomes unity whatever the
astrophysical parameters.

Inspecting directly the positron fluxes, typically,  for a 100 GeV DM
particle annihilating into a $\bar{b} b $ pair, uncertainties due to propagation on the positron
flux are one order of magnitude at 1 GeV and a factor of two  at 10 GeV and above.
We find an increasing uncertainty for harder source spectra, heavier  DM, steeper profiles.

The comparison with current data shows that  the possible HEAT excess is reproduced  for DM
annihilating mostly into gauge bosons or directly into a positron--electron pair, and the
agreement is not limited by the astrophysical uncertainties. A boost factor of 10 is enough to
obtain an excellent agreement between the measurements and the median flux, for a 100 GeV DM
particle. A smaller value would be required for a flux at the upper envelope of the uncertainty
band.

We have finally drawn prospects for two interesting 3--year flight
space missions, like PAMELA, already in operation, and the future
AMS-02. We reach the remarkable conclusion that not only will
PAMELA have the capability to disentangle the signal from the
background, but it will also distinguish among different astrophysical
models, especially for hard spectra. For AMS-02 the possibility to
disentangle the signal from the background is also clearly
manifest. We also wish to remind that improved experimental results on cosmic ray nuclei,
expecially on the B/C ratio, will be instrumental to improve the determination 
of the parameters of the propagation models, and will therefore lead to sharper theoretical 
predictions. This in turn will lead to a more refined comparison with the
experimental data on the positron flux. Moreover, a good
determination of the unstable/stable nuclei abundances like
the $^{10}$Be/$^{9}$Be ratio could shed some light on the local environment, which is
certainly mostly relevant to the positrons.

In the present paper we have thus presented the methods and the practical tools to evaluate
the primary positron fluxes in detailed propagation models.
We have provided careful estimations of the underlying uncertainties and shown the extraordinary
potentials of already running, or near to come, space detectors.

\acknowledgments
R.L., F.D. and N.F gratefully acknowledge financial support
provided by Research Grants of the Italian Ministero
dell'Istruzione, dell'Universit\`a e della Ricerca (MIUR), of the
Universit\`a di Torino and of the Istituto Nazionale di Fisica
Nucleare (INFN) within the {\sl Astroparticle Physics Project}.
R.L. also acknowledges the Comisi\'on Nacional de Investigaci\'on
Cient\'ifica y Tecnol\'ogica (CONICYT) of Chile. T.D. acknowledges
financial support from the French \'Ecole Polytechnique and P.S.
is grateful to the French Programme National de Cosmologie.

%
\bibliography{draft}

\begin{thebibliography}{27}
\expandafter\ifx\csname natexlab\endcsname\relax\def\natexlab#1{#1}\fi
\expandafter\ifx\csname bibnamefont\endcsname\relax
  \def\bibnamefont#1{#1}\fi
\expandafter\ifx\csname bibfnamefont\endcsname\relax
  \def\bibfnamefont#1{#1}\fi
\expandafter\ifx\csname citenamefont\endcsname\relax
  \def\citenamefont#1{#1}\fi
\expandafter\ifx\csname url\endcsname\relax
  \def\url#1{\texttt{#1}}\fi
\expandafter\ifx\csname urlprefix\endcsname\relax\def\urlprefix{URL }\fi
\providecommand{\bibinfo}[2]{#2}
\providecommand{\eprint}[2][]{\url{#2}}

\bibitem[{\citenamefont{Moskalenko and Strong}(1998)}]{strongmoska}
\bibinfo{author}{\bibfnamefont{I.~V.} \bibnamefont{Moskalenko}}
  \bibnamefont{and} \bibinfo{author}{\bibfnamefont{A.~W.}
  \bibnamefont{Strong}}, \bibinfo{journal}{Astrophys. J.}
  \textbf{\bibinfo{volume}{493}}, \bibinfo{pages}{694} (\bibinfo{year}{1998}),
  \eprint{astro-ph/9710124}.

\bibitem[{\citenamefont{Barwick et~al.}(1997)}]{heat}
\bibinfo{author}{\bibfnamefont{S.~W.} \bibnamefont{Barwick}}
  \bibnamefont{et~al.} (\bibinfo{collaboration}{HEAT}),
  \bibinfo{journal}{Astrophys. J.} \textbf{\bibinfo{volume}{482}},
  \bibinfo{pages}{L191} (\bibinfo{year}{1997}), \eprint{astro-ph/9703192}.

\bibitem[{\citenamefont{Ahlen et~al.}(1994)}]{ams}
\bibinfo{author}{\bibfnamefont{S.}~\bibnamefont{Ahlen}} \bibnamefont{et~al.},
  \bibinfo{journal}{Nucl. Instrum. Meth. A} \textbf{\bibinfo{volume}{350}},
  \bibinfo{pages}{351} (\bibinfo{year}{1994}).

\bibitem[{\citenamefont{Alcaraz et~al.}(2000)}]{ams01}
\bibinfo{author}{\bibfnamefont{J.}~\bibnamefont{Alcaraz}} \bibnamefont{et~al.}
  (\bibinfo{collaboration}{AMS}), \bibinfo{journal}{Phys. Lett.}
  \textbf{\bibinfo{volume}{B484}}, \bibinfo{pages}{10} (\bibinfo{year}{2000}).

\bibitem[{\citenamefont{Aguilar et~al.}(2007)}]{ams02}
\bibinfo{author}{\bibfnamefont{M.}~\bibnamefont{Aguilar}} \bibnamefont{et~al.}
  (\bibinfo{collaboration}{AMS-01}), \bibinfo{journal}{Phys. Lett.}
  \textbf{\bibinfo{volume}{B646}}, \bibinfo{pages}{145} (\bibinfo{year}{2007}),
  \eprint{astro-ph/0703154}.

\bibitem[{\citenamefont{{Boezio} et~al.}(2000)\citenamefont{{Boezio},
  {Carlson}, {Francke}, {Weber}, {Suffert}, {Hof}, {Menn}, {Simon}, {Stephens},
  {Bellotti} et~al.}}]{2000ApJ...532..653B}
\bibinfo{author}{\bibfnamefont{M.}~\bibnamefont{{Boezio}}},
  \bibinfo{author}{\bibfnamefont{P.}~\bibnamefont{{Carlson}}},
  \bibinfo{author}{\bibfnamefont{T.}~\bibnamefont{{Francke}}},
  \bibinfo{author}{\bibfnamefont{N.}~\bibnamefont{{Weber}}},
  \bibinfo{author}{\bibfnamefont{M.}~\bibnamefont{{Suffert}}},
  \bibinfo{author}{\bibfnamefont{M.}~\bibnamefont{{Hof}}},
  \bibinfo{author}{\bibfnamefont{W.}~\bibnamefont{{Menn}}},
  \bibinfo{author}{\bibfnamefont{M.}~\bibnamefont{{Simon}}},
  \bibinfo{author}{\bibfnamefont{S.~A.} \bibnamefont{{Stephens}}},
  \bibinfo{author}{\bibfnamefont{R.}~\bibnamefont{{Bellotti}}},
  \bibnamefont{et~al.}, \bibinfo{journal}{Astrophys. J.}
  \textbf{\bibinfo{volume}{532}}, \bibinfo{pages}{653} (\bibinfo{year}{2000}).

\bibitem[{\citenamefont{{Grimani} et~al.}(2002)\citenamefont{{Grimani},
  {Stephens}, {Cafagna}, {Basini}, {Bellotti}, {Brunetti}, {Circella},
  {Codino}, {De Marzo}, {De Pascale} et~al.}}]{2002A&A...392..287G}
\bibinfo{author}{\bibfnamefont{C.}~\bibnamefont{{Grimani}}},
  \bibinfo{author}{\bibfnamefont{S.~A.} \bibnamefont{{Stephens}}},
  \bibinfo{author}{\bibfnamefont{F.~S.} \bibnamefont{{Cafagna}}},
  \bibinfo{author}{\bibfnamefont{G.}~\bibnamefont{{Basini}}},
  \bibinfo{author}{\bibfnamefont{R.}~\bibnamefont{{Bellotti}}},
  \bibinfo{author}{\bibfnamefont{M.~T.} \bibnamefont{{Brunetti}}},
  \bibinfo{author}{\bibfnamefont{M.}~\bibnamefont{{Circella}}},
  \bibinfo{author}{\bibfnamefont{A.}~\bibnamefont{{Codino}}},
  \bibinfo{author}{\bibfnamefont{C.}~\bibnamefont{{De Marzo}}},
  \bibinfo{author}{\bibfnamefont{M.~P.} \bibnamefont{{De Pascale}}},
  \bibnamefont{et~al.}, \bibinfo{journal}{Astronomy \& Astrophysics}
  \textbf{\bibinfo{volume}{392}}, \bibinfo{pages}{287} (\bibinfo{year}{2002}).

\bibitem[{\citenamefont{Baltz and Edsjo}(1999)}]{baltz_edsjo99}
\bibinfo{author}{\bibfnamefont{E.~A.} \bibnamefont{Baltz}} \bibnamefont{and}
  \bibinfo{author}{\bibfnamefont{J.}~\bibnamefont{Edsjo}},
  \bibinfo{journal}{Phys. Rev.} \textbf{\bibinfo{volume}{D59}},
  \bibinfo{pages}{023511} (\bibinfo{year}{1999}), \eprint{astro-ph/9808243}.

\bibitem[{\citenamefont{Hooper and Silk}(2005)}]{Hooper:2004bq}
\bibinfo{author}{\bibfnamefont{D.}~\bibnamefont{Hooper}} \bibnamefont{and}
  \bibinfo{author}{\bibfnamefont{J.}~\bibnamefont{Silk}},
  \bibinfo{journal}{Phys. Rev.} \textbf{\bibinfo{volume}{D71}},
  \bibinfo{pages}{083503} (\bibinfo{year}{2005}), \eprint{hep-ph/0409104}.

\bibitem[{\citenamefont{Lavalle
  et~al.}(2007{\natexlab{a}})\citenamefont{Lavalle, Pochon, Salati, and
  Taillet}}]{Lavalle:2006vb}
\bibinfo{author}{\bibfnamefont{J.}~\bibnamefont{Lavalle}},
  \bibinfo{author}{\bibfnamefont{J.}~\bibnamefont{Pochon}},
  \bibinfo{author}{\bibfnamefont{P.}~\bibnamefont{Salati}}, \bibnamefont{and}
  \bibinfo{author}{\bibfnamefont{R.}~\bibnamefont{Taillet}},
  \bibinfo{journal}{Astronomy \& Astrophysics} \textbf{\bibinfo{volume}{462}},
  \bibinfo{pages}{827} (\bibinfo{year}{2007}{\natexlab{a}}),
  \eprint{astro-ph/0603796}.

\bibitem[{\citenamefont{Lavalle
  et~al.}(2007{\natexlab{b}})\citenamefont{Lavalle, Yuan, Maurin, and
  Bi}}]{Lavalle:1900wn}
\bibinfo{author}{\bibfnamefont{J.}~\bibnamefont{Lavalle}},
  \bibinfo{author}{\bibfnamefont{Q.}~\bibnamefont{Yuan}},
  \bibinfo{author}{\bibfnamefont{D.}~\bibnamefont{Maurin}}, \bibnamefont{and}
  \bibinfo{author}{\bibfnamefont{X.~J.} \bibnamefont{Bi}}
  (\bibinfo{year}{2007}{\natexlab{b}}), \eprint{arXiv:0709.3634 [astro-ph]}.

\bibitem[{\citenamefont{Maurin et~al.}(2001)\citenamefont{Maurin, Donato,
  Taillet, and Salati}}]{parfit}
\bibinfo{author}{\bibfnamefont{D.}~\bibnamefont{Maurin}},
  \bibinfo{author}{\bibfnamefont{F.}~\bibnamefont{Donato}},
  \bibinfo{author}{\bibfnamefont{R.}~\bibnamefont{Taillet}}, \bibnamefont{and}
  \bibinfo{author}{\bibfnamefont{P.}~\bibnamefont{Salati}},
  \bibinfo{journal}{Astrophys. J.} \textbf{\bibinfo{volume}{555}},
  \bibinfo{pages}{585} (\bibinfo{year}{2001}).

\bibitem[{\citenamefont{Cheng et~al.}(2002)\citenamefont{Cheng, Matchev, and
  Schmaltz}}]{LKP_model_1}
\bibinfo{author}{\bibfnamefont{H.-C.} \bibnamefont{Cheng}},
  \bibinfo{author}{\bibfnamefont{K.~T.} \bibnamefont{Matchev}},
  \bibnamefont{and} \bibinfo{author}{\bibfnamefont{M.}~\bibnamefont{Schmaltz}},
  \bibinfo{journal}{Phys. Rev.} \textbf{\bibinfo{volume}{D66}},
  \bibinfo{pages}{036005} (\bibinfo{year}{2002}), \eprint{hep-ph/0204342}.

\bibitem[{\citenamefont{Servant and Tait}(2003)}]{LKP_model_2}
\bibinfo{author}{\bibfnamefont{G.}~\bibnamefont{Servant}} \bibnamefont{and}
  \bibinfo{author}{\bibfnamefont{T.~M.~P.} \bibnamefont{Tait}},
  \bibinfo{journal}{Nucl. Phys.} \textbf{\bibinfo{volume}{B650}},
  \bibinfo{pages}{391} (\bibinfo{year}{2003}), \eprint{hep-ph/0206071}.

\bibitem[{\citenamefont{Appelquist et~al.}(2001)\citenamefont{Appelquist,
  Cheng, and Dobrescu}}]{UED_models}
\bibinfo{author}{\bibfnamefont{T.}~\bibnamefont{Appelquist}},
  \bibinfo{author}{\bibfnamefont{H.-C.} \bibnamefont{Cheng}}, \bibnamefont{and}
  \bibinfo{author}{\bibfnamefont{B.~A.} \bibnamefont{Dobrescu}},
  \bibinfo{journal}{Phys. Rev.} \textbf{\bibinfo{volume}{D64}},
  \bibinfo{pages}{035002} (\bibinfo{year}{2001}), \eprint{hep-ph/0012100}.

\bibitem[{\citenamefont{Bottino
  et~al.}(2003{\natexlab{a}})\citenamefont{Bottino, Fornengo, and
  Scopel}}]{Bottino:2002ry}
\bibinfo{author}{\bibfnamefont{A.}~\bibnamefont{Bottino}},
  \bibinfo{author}{\bibfnamefont{N.}~\bibnamefont{Fornengo}}, \bibnamefont{and}
  \bibinfo{author}{\bibfnamefont{S.}~\bibnamefont{Scopel}},
  \bibinfo{journal}{Phys. Rev.} \textbf{\bibinfo{volume}{D67}},
  \bibinfo{pages}{063519} (\bibinfo{year}{2003}{\natexlab{a}}),
  \eprint{hep-ph/0212379}.

\bibitem[{\citenamefont{Bottino
  et~al.}(2003{\natexlab{b}})\citenamefont{Bottino, Donato, Fornengo, and
  Scopel}}]{Bottino:2003iu}
\bibinfo{author}{\bibfnamefont{A.}~\bibnamefont{Bottino}},
  \bibinfo{author}{\bibfnamefont{F.}~\bibnamefont{Donato}},
  \bibinfo{author}{\bibfnamefont{N.}~\bibnamefont{Fornengo}}, \bibnamefont{and}
  \bibinfo{author}{\bibfnamefont{S.}~\bibnamefont{Scopel}},
  \bibinfo{journal}{Phys. Rev.} \textbf{\bibinfo{volume}{D68}},
  \bibinfo{pages}{043506} (\bibinfo{year}{2003}{\natexlab{b}}),
  \eprint{hep-ph/0304080}.

\bibitem[{\citenamefont{Belanger et~al.}(2003)\citenamefont{Belanger, Boudjema,
  Pukhov, and Rosier-Lees}}]{Belanger:2002nr}
\bibinfo{author}{\bibfnamefont{G.}~\bibnamefont{Belanger}},
  \bibinfo{author}{\bibfnamefont{F.}~\bibnamefont{Boudjema}},
  \bibinfo{author}{\bibfnamefont{A.}~\bibnamefont{Pukhov}}, \bibnamefont{and}
  \bibinfo{author}{\bibfnamefont{S.}~\bibnamefont{Rosier-Lees}},
  \bibinfo{journal}{Phys. Rev.} \textbf{\bibinfo{volume}{D68}}
  (\bibinfo{year}{2003}), \eprint{hep-ph/0310037}.

\bibitem[{\citenamefont{Hooper and Plehn}(2003)}]{Hooper:2002nq}
\bibinfo{author}{\bibfnamefont{D.}~\bibnamefont{Hooper}} \bibnamefont{and}
  \bibinfo{author}{\bibfnamefont{T.}~\bibnamefont{Plehn}},
  \bibinfo{journal}{Phys. Lett.} \textbf{\bibinfo{volume}{B562}},
  \bibinfo{pages}{18} (\bibinfo{year}{2003}), \eprint{hep-ph/0212226}.

\bibitem[{\citenamefont{Sjostrand et~al.}(2001)}]{Pythia}
\bibinfo{author}{\bibfnamefont{T.}~\bibnamefont{Sjostrand}}
  \bibnamefont{et~al.}, \bibinfo{journal}{Comput. Phys. Commun.}
  \textbf{\bibinfo{volume}{135}}, \bibinfo{pages}{238} (\bibinfo{year}{2001}),
  \eprint{hep-ph/0010017}.

\bibitem[{\citenamefont{Bahcall and Soneira}(1980)}]{bahcall}
\bibinfo{author}{\bibfnamefont{J.~N.} \bibnamefont{Bahcall}} \bibnamefont{and}
  \bibinfo{author}{\bibfnamefont{R.~M.} \bibnamefont{Soneira}},
  \bibinfo{journal}{Astrophys. J. Suppl.} \textbf{\bibinfo{volume}{44}},
  \bibinfo{pages}{73} (\bibinfo{year}{1980}).

\bibitem[{\citenamefont{Navarro et~al.}(1997)\citenamefont{Navarro, Frenk, and
  White}}]{nfw}
\bibinfo{author}{\bibfnamefont{J.~F.} \bibnamefont{Navarro}},
  \bibinfo{author}{\bibfnamefont{C.~S.} \bibnamefont{Frenk}}, \bibnamefont{and}
  \bibinfo{author}{\bibfnamefont{S.~D.~M.} \bibnamefont{White}},
  \bibinfo{journal}{Astrophys. J.} \textbf{\bibinfo{volume}{490}},
  \bibinfo{pages}{493} (\bibinfo{year}{1997}), \eprint{astro-ph/9611107}.

\bibitem[{\citenamefont{Diemand et~al.}(2004)\citenamefont{Diemand, Moore, and
  Stadel}}]{moore}
\bibinfo{author}{\bibfnamefont{J.}~\bibnamefont{Diemand}},
  \bibinfo{author}{\bibfnamefont{B.}~\bibnamefont{Moore}}, \bibnamefont{and}
  \bibinfo{author}{\bibfnamefont{J.}~\bibnamefont{Stadel}},
  \bibinfo{journal}{Mon. Not. Roy. Astron. Soc.}
  \textbf{\bibinfo{volume}{353}}, \bibinfo{pages}{624} (\bibinfo{year}{2004}),
  \eprint{astro-ph/0402267}.

\bibitem[{\citenamefont{Donato et~al.}(2004)\citenamefont{Donato, Fornengo,
  Maurin, Salati, and Taillet}}]{primary_pbar}
\bibinfo{author}{\bibfnamefont{F.}~\bibnamefont{Donato}},
  \bibinfo{author}{\bibfnamefont{N.}~\bibnamefont{Fornengo}},
  \bibinfo{author}{\bibfnamefont{D.}~\bibnamefont{Maurin}},
  \bibinfo{author}{\bibfnamefont{P.}~\bibnamefont{Salati}}, \bibnamefont{and}
  \bibinfo{author}{\bibfnamefont{R.}~\bibnamefont{Taillet}},
  \bibinfo{journal}{Phys. Rev.} \textbf{\bibinfo{volume}{D69}},
  \bibinfo{pages}{063501} (\bibinfo{year}{2004}).

\bibitem[{\citenamefont{Donato et~al.}(2001)}]{secondary_pbar}
\bibinfo{author}{\bibfnamefont{F.}~\bibnamefont{Donato}} \bibnamefont{et~al.},
  \bibinfo{journal}{Astrophys. J.} \textbf{\bibinfo{volume}{563}},
  \bibinfo{pages}{172} (\bibinfo{year}{2001}).

\bibitem[{\citenamefont{Maurin and Taillet}(2003)}]{Maurin:2002uc}
\bibinfo{author}{\bibfnamefont{D.}~\bibnamefont{Maurin}} \bibnamefont{and}
  \bibinfo{author}{\bibfnamefont{R.}~\bibnamefont{Taillet}},
  \bibinfo{journal}{Astron. Astrophys.} \textbf{\bibinfo{volume}{404}},
  \bibinfo{pages}{949} (\bibinfo{year}{2003}), \eprint{astro-ph/0212113}.

\bibitem[{\citenamefont{Boezio et~al.}(2004)}]{Boezio:2004jx}
\bibinfo{author}{\bibfnamefont{M.}~\bibnamefont{Boezio}} \bibnamefont{et~al.},
  \bibinfo{journal}{Nucl. Phys. Proc. Suppl.} \textbf{\bibinfo{volume}{134}},
  \bibinfo{pages}{39} (\bibinfo{year}{2004}).

\end{thebibliography}

\end{document}